\pdfoutput=1
\documentclass[12pt,a4paper]{article}

\usepackage{ifthen} 
\newboolean{pdflatex}
\setboolean{pdflatex}{true} 

\newboolean{articletitles}
\setboolean{articletitles}{true} 

\newboolean{uprightparticles}
\setboolean{uprightparticles}{false} 

\def\paperauthors{LHCb collaboration} 
\def\paperasciititle{Measurement of CP asymmetry in BsDsK decays} 
\def\papertitle{Measurement of $\CP$ asymmetry in \BsDsK decays} 
\def\paperkeywords{{High Energy Physics}, {LHCb}} 
\def\papercopyright{\the\year\ CERN for the benefit of the LHCb collaboration} 
\def\paperlicence{CC BY 4.0 licence}
\def\paperlicenceurl{https://creativecommons.org/licenses/by/4.0/}

\usepackage[top=1in, bottom=1.25in, left=1in, right=1in]{geometry}

\usepackage{microtype}
\usepackage{lineno}  
\usepackage{xspace} 
\usepackage{caption} 

\usepackage{graphicx}  
\usepackage{color}
\usepackage{colortbl}
\graphicspath{{./figs/}} 

\usepackage{amsmath} 
\usepackage{amssymb}
\usepackage{amsfonts}
\usepackage{upgreek} 

\newcommand*\patchAmsMathEnvironmentForLineno[1]{%
\expandafter\let\csname old#1\expandafter\endcsname\csname #1\endcsname
\expandafter\let\csname oldend#1\expandafter\endcsname\csname
end#1\endcsname
 \renewenvironment{#1}%
   {\linenomath\csname old#1\endcsname}%
   {\csname oldend#1\endcsname\endlinenomath}%
}
\newcommand*\patchBothAmsMathEnvironmentsForLineno[1]{%
  \patchAmsMathEnvironmentForLineno{#1}%
  \patchAmsMathEnvironmentForLineno{#1*}%
}
\AtBeginDocument{%
\patchBothAmsMathEnvironmentsForLineno{equation}%
\patchBothAmsMathEnvironmentsForLineno{align}%
\patchBothAmsMathEnvironmentsForLineno{flalign}%
\patchBothAmsMathEnvironmentsForLineno{alignat}%
\patchBothAmsMathEnvironmentsForLineno{gather}%
\patchBothAmsMathEnvironmentsForLineno{multline}%
\patchBothAmsMathEnvironmentsForLineno{eqnarray}%
}

\usepackage{hyperxmp}

\usepackage[pdftex,
            pdfauthor={\paperauthors},
            pdftitle={\paperasciititle},
            pdfkeywords={\paperkeywords},
            pdfcopyright={Copyright (C) \papercopyright},
            pdflicenseurl={\paperlicenceurl}]{hyperref}

\usepackage[colorinlistoftodos,textsize=scriptsize]{todonotes}

\usepackage[bottom,flushmargin,hang,multiple]{footmisc}

\usepackage[all]{hypcap} 

\usepackage{xspace} 
\usepackage{upgreek}

\def\lhcb   {\mbox{LHCb}\xspace}

\def\babar  {\mbox{BaBar}\xspace}
\def\belle  {\mbox{Belle}\xspace}

\def\MagUp {\mbox{\em Mag\kern -0.05em Up}\xspace}

\ifthenelse{\boolean{uprightparticles}}%
{
 
 \def\Pgamma      {\ensuremath{\upgamma}\xspace}

 \def\Ppi         {\ensuremath{\uppi}\xspace}                 
                  
 \def\Prho        {\ensuremath{\uprho}\xspace}

 \def\Ppsi        {\ensuremath{\uppsi}\xspace}

 \def\PDelta      {\ensuremath{\Delta}\xspace}                 
 \def\PXi         {\ensuremath{\Xi}\xspace}                 
 \def\PLambda     {\ensuremath{\Lambda}\xspace}                 
 \def\PSigma      {\ensuremath{\Sigma}\xspace}                 
 \def\POmega      {\ensuremath{\Omega}\xspace}                 
 \def\PUpsilon    {\ensuremath{\Upsilon}\xspace}
 \let\oldPi\Pi
 \def\PPi         {\ensuremath{\oldPi}\xspace}

 \def\PB      {\ensuremath{\mathrm{B}}\xspace}                 
                  
 \def\PD      {\ensuremath{\mathrm{D}}\xspace}

 \def\PJ      {\ensuremath{\mathrm{J}}\xspace}                 
 \def\PK      {\ensuremath{\mathrm{K}}\xspace}

 \def\Pb      {\ensuremath{\mathrm{b}}\xspace}                 
 \def\Pc      {\ensuremath{\mathrm{c}}\xspace}

 \def\Pi      {\ensuremath{\mathrm{i}}\xspace}

 \def\Pp      {\ensuremath{\mathrm{p}}\xspace}

 \def\Ps      {\ensuremath{\mathrm{s}}\xspace}

 \def\thebaroffset{0.0em}
}
{
 
 \def\Pgamma      {\ensuremath{\gamma}\xspace}

 \def\Ppi         {\ensuremath{\pi}\xspace}                 
                  
 \def\Prho        {\ensuremath{\rho}\xspace}

 \def\Ppsi        {\ensuremath{\psi}\xspace}                 
                  
 \mathchardef\PDelta="7101
 \mathchardef\PXi="7104
 \mathchardef\PLambda="7103
 \mathchardef\PSigma="7106
 \mathchardef\POmega="710A
 \mathchardef\PUpsilon="7107
 \mathchardef\PPi="7105
                  
 \def\PB      {\ensuremath{B}\xspace}                 
                  
 \def\PD      {\ensuremath{D}\xspace}

 \def\PJ      {\ensuremath{J}\xspace}                 
 \def\PK      {\ensuremath{K}\xspace}

 \def\Pb      {\ensuremath{b}\xspace}                 
 \def\Pc      {\ensuremath{c}\xspace}

 \def\Pi      {\ensuremath{i}\xspace}

 \def\Pp      {\ensuremath{p}\xspace}

 \def\Ps      {\ensuremath{s}\xspace}

 \def\thebaroffset{0.18em}
}
\newcommand{\offsetoverline}[2][\thebaroffset]{\kern #1\overline{\kern -#1 #2}}%

\makeatletter
\ifcase \@ptsize \relax
  \newcommand{\miniscule}{\@setfontsize\miniscule{4}{5}}
\or
  \newcommand{\miniscule}{\@setfontsize\miniscule{5}{6}}
\or
  \newcommand{\miniscule}{\@setfontsize\miniscule{5}{6}}
\fi
\makeatother

\DeclareRobustCommand{\optbar}[1]{\shortstack{{\miniscule (\rule[.5ex]{1.25em}{.18mm})}
  \\ [-.7ex] $#1$}}

\def\g      {{\ensuremath{\Pgamma}}\xspace}

\def\squark    {{\ensuremath{\Ps}}\xspace}
\def\squarkbar {{\ensuremath{\overline \squark}}\xspace}

\def\cquark    {{\ensuremath{\Pc}}\xspace}

\def\bquark    {{\ensuremath{\Pb}}\xspace}
\def\bquarkbar {{\ensuremath{\overline \bquark}}\xspace}

\def\pion   {{\ensuremath{\Ppi}}\xspace}
\def\piz    {{\ensuremath{\pion^0}}\xspace}
\def\pip    {{\ensuremath{\pion^+}}\xspace}
\def\pim    {{\ensuremath{\pion^-}}\xspace}

\def\rhomeson {{\ensuremath{\Prho}}\xspace}

\def\rhop     {{\ensuremath{\rhomeson^+}}\xspace}

\def\kaon    {{\ensuremath{\PK}}\xspace}
\def\Kbar    {{\ensuremath{\offsetoverline{\PK}}}\xspace}

\def\KorKbar {\kern \thebaroffset\optbar{\kern -\thebaroffset \PK}{}\xspace}

\def\Kzb     {{\ensuremath{\Kbar{}^0}}\xspace}
\def\Kp      {{\ensuremath{\kaon^+}}\xspace}
\def\Km      {{\ensuremath{\kaon^-}}\xspace}

\def\Kstarz  {{\ensuremath{\kaon^{*0}}}\xspace}

\def\D       {{\ensuremath{\PD}}\xspace}

\def\DorDbar {\kern \thebaroffset\optbar{\kern -\thebaroffset \PD}\xspace}
\def\Dz      {{\ensuremath{\D^0}}\xspace}

\def\Dp      {{\ensuremath{\D^+}}\xspace}
\def\Dm      {{\ensuremath{\D^-}}\xspace}

\def\DpDm    {\ensuremath{\Dp {\kern -0.16em \Dm}}\xspace}

\def\Dstarp  {{\ensuremath{\D^{*+}}}\xspace}

\def\Ds      {{\ensuremath{\D^+_\squark}}\xspace}
\def\Dsp     {{\ensuremath{\D^+_\squark}}\xspace}
\def\Dsm     {{\ensuremath{\D^-_\squark}}\xspace}

\def\Dsmp    {{\ensuremath{\D^{\mp}_\squark}}\xspace}

\def\Dssm    {{\ensuremath{\D^{*-}_\squark}}\xspace}

\def\B       {{\ensuremath{\PB}}\xspace}
\def\Bbar    {{\ensuremath{\offsetoverline{\PB}}}\xspace}

\def\BorBbar {\kern \thebaroffset\optbar{\kern -\thebaroffset \PB}\xspace}
\def\Bz      {{\ensuremath{\B^0}}\xspace}

\def\Bd      {{\ensuremath{\B^0}}\xspace}
\def\Bdb     {{\ensuremath{\Bbar{}^0}}\xspace}
\def\BdorBdbar {\kern \thebaroffset\optbar{\kern -\thebaroffset \Bd}\xspace}
\def\Bu      {{\ensuremath{\B^+}}\xspace}

\def\Bp      {{\ensuremath{\Bu}}\xspace}

\def\Bs      {{\ensuremath{\B^0_\squark}}\xspace}
\def\Bsb     {{\ensuremath{\Bbar{}^0_\squark}}\xspace}
\def\BsorBsbar {\kern \thebaroffset\optbar{\kern -\thebaroffset \Bs}\xspace}

\def\jpsi     {{\ensuremath{{\PJ\mskip -3mu/\mskip -2mu\Ppsi}}}\xspace}

\def\Y#1S{\ensuremath{\PUpsilon{(#1S)}}\xspace}

\def\proton      {{\ensuremath{\Pp}}\xspace}

\def\Lz          {{\ensuremath{\PLambda}}\xspace}
\def\Lbar        {{\ensuremath{\offsetoverline{\PLambda}}}\xspace}
\def\LorLbar     {\kern \thebaroffset\optbar{\kern -\thebaroffset \PLambda}\xspace}

\def\Lcbar       {{\ensuremath{\Lbar{}^-_\cquark}}\xspace}

\def\Lb           {{\ensuremath{\Lz^0_\bquark}}\xspace}
\def\Lbbar        {{\ensuremath{\Lbar{}^0_\bquark}}\xspace}

\newcommand{\decay}[2]{\ensuremath{#1\!\to #2}\xspace} 

\def\to                 {\ensuremath{\rightarrow}\xspace}

\def\CP                {{\ensuremath{C\!P}}\xspace}

\newcommand{\dms}{{\ensuremath{\Delta m_{\squark}}}\xspace}

\newcommand{\DGs}{{\ensuremath{\Delta\Gamma_{\squark}}}\xspace}

\newcommand{\Gs}{{\ensuremath{\Gamma_{\squark}}}\xspace}

\newcommand{\phis}{{\ensuremath{\phi_{\squark}}}\xspace}
\newcommand{\betas}{{\ensuremath{\beta_{\squark}}}\xspace}

\def\AT#1     {\ensuremath{A_{\mathrm{T}}^{#1}}\xspace}           

\def\C#1      {\ensuremath{\mathcal{C}_{#1}}\xspace}                       
\def\Cp#1     {\ensuremath{\mathcal{C}_{#1}^{'}}\xspace}                    
\def\Ceff#1   {\ensuremath{\mathcal{C}_{#1}^{\mathrm{(eff)}}}\xspace}        
\def\Cpeff#1  {\ensuremath{\mathcal{C}_{#1}^{'\mathrm{(eff)}}}\xspace}       
\def\Ope#1    {\ensuremath{\mathcal{O}_{#1}}\xspace}                       
\def\Opep#1   {\ensuremath{\mathcal{O}_{#1}^{'}}\xspace}                    

\newcommand{\nospaceunit}[1]{\ensuremath{\text{#1}}}       
\newcommand{\aunit}[1]{\ensuremath{\text{\,#1}}}       

\newcommand{\tev}{\aunit{Te\kern -0.1em V}\xspace}
\newcommand{\gev}{\aunit{Ge\kern -0.1em V}\xspace}
\newcommand{\mev}{\aunit{Me\kern -0.1em V}\xspace}
\newcommand{\kev}{\aunit{ke\kern -0.1em V}\xspace}
\newcommand{\ev}{\aunit{e\kern -0.1em V}\xspace}
 
\newcommand{\mevc}{\ensuremath{\aunit{Me\kern -0.1em V\!/}c}\xspace}
\newcommand{\gevc}{\ensuremath{\aunit{Ge\kern -0.1em V\!/}c}\xspace}
\newcommand{\mevcc}{\ensuremath{\aunit{Me\kern -0.1em V\!/}c^2}\xspace}
\newcommand{\gevcc}{\ensuremath{\aunit{Ge\kern -0.1em V\!/}c^2}\xspace}

\def\mum  {\ensuremath{\,\upmu\nospaceunit{m}}\xspace}

\def\fb   {\ensuremath{\aunit{fb}}\xspace}
\def\invfb   {\ensuremath{\fb^{-1}}\xspace}

\def\ps   {\ensuremath{\aunit{ps}}\xspace}
\def\fs   {\aunit{fs}}

\def\invps{\ensuremath{\ps^{-1}}\xspace}

\newcommand{\chisqip}{\ensuremath{\chi^2_{\text{IP}}}\xspace}

\def\gsim{{~\raise.15em\hbox{$>$}\kern-.85em
          \lower.35em\hbox{$\sim$}~}\xspace}
\def\lsim{{~\raise.15em\hbox{$<$}\kern-.85em
          \lower.35em\hbox{$\sim$}~}\xspace}

\newcommand{\Real}{\ensuremath{\mathcal{R}e}\xspace}
\newcommand{\Imag}{\ensuremath{\mathcal{I}m}\xspace}

\def\sqs   {\ensuremath{\protect\sqrt{s}}\xspace}

\def\pt         {\ensuremath{p_{\mathrm{T}}}\xspace}

\def\ptot       {\ensuremath{p}\xspace}

\def\rad{\aunit{rad}\xspace}

\def\evtgen     {\mbox{\textsc{EvtGen}}\xspace}

\def\geant      {\mbox{\textsc{Geant4}}\xspace}

\def\photos     {\mbox{\textsc{Photos}}\xspace}

\def\pythia     {\mbox{\textsc{Pythia}}\xspace}

\def\tell1  {TELL1\xspace}
\def\ukl1   {UKL1\xspace}

\newcommand{\ie}{\mbox{\itshape i.e.}\xspace}

\newcommand{\vs}{\mbox{\itshape vs.}\xspace}

\newcommand{\lhcborcid}[1]{\href{https://orcid.org/#1}{\hspace*{0.1em}\raisebox{-0.45ex}{\includegraphics[width=1em]{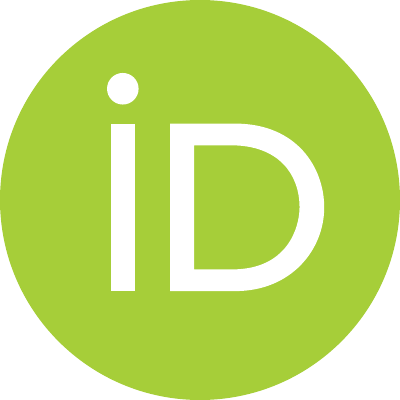}}}}

\def\f{f}

\newcommand{\weak}{\ensuremath{\gamma - 2\betas}\xspace}
\newcommand{\rdsk}      {\ensuremath{r_{D_sK}}\xspace}
\newcommand{\strong}    {\ensuremath{\delta}\xspace}
\newcommand{\omcl}      {\ensuremath{1-{\rm CL}}\xspace}

\def\Af      {{\ensuremath{A_f}}\xspace}

\def\Abf     {{\ensuremath{\bar{A}_f}}\xspace}

\def\lf      {{\ensuremath{\lambda_f}}\xspace}
\def\lfb     {{\ensuremath{\lambda_{\bar{f}}}}\xspace}

\def\Cf      {{\ensuremath{C_f}}\xspace}
\def\AdGamf  {{\ensuremath{A^{\Delta \Gamma}_f}}\xspace}
\def\Sf      {{\ensuremath{S_f}}\xspace}
\def\Cfb     {{\ensuremath{C_{\bar{f}}}}\xspace}
\def\AdGamfb {{\ensuremath{A^{\Delta \Gamma}_{\bar{f}}}}\xspace}
\def\Sfb     {{\ensuremath{S_{\bar{f}}}}\xspace}

\renewcommand{\Ds}       {\texorpdfstring{\ensuremath{D_s^-}}{Ds}\xspace}

\newcommand{\BdDK}     {\texorpdfstring{\decay{\Bz}{\Dm \Kp}}{}}
\newcommand{\BdDPi}    {\texorpdfstring{\decay{\Bz}{\Dm \pip}}{}}
\newcommand{\BdDh}     {\decay{\Bz}{\Dm h^{+}}}

\newcommand{\BdDsK}    {\texorpdfstring{\decay{\Bz}{\Dsm \Kp}}{}}

\newcommand{\BsDsK}    {\texorpdfstring{\decay{B^0_s}{D_s^\mp K^\pm}}{}}
\newcommand{\BsDsPi}   {\texorpdfstring{\decay{B^0_s}{D_s^- \pi^+}}{}}

\newcommand{\BsDsRho}  {\decay{\Bs}{\Dsm \rho^{+}}}

\newcommand{\BsDspKm}  {\texorpdfstring{\decay{\Bs}{D_s^+ K^-}}{}}
\newcommand{\BsDsmKp}  {\texorpdfstring{\decay{\Bs}{D_s^- K^+}}{}}

\newcommand{\BsDsstPi} {\decay{\Bs}{\Dssm \pip}}
\newcommand{\DKPiPi}   {\decay{\Dm}{\Kp\pim\pim}}
\newcommand{\DsKKPi}   {\decay{\Dsm}{\Km\Kp\pim}}
\newcommand{\DsKPiPi}  {\decay{\Dsm}{\Km\pip\pim}}

\newcommand{\DsK}      {\texorpdfstring{D_s^\mp K^\pm}{}}

\newcommand{\DsNonRes} {\decay{\Dsm}{(K^-K^+\pi^-)_{\rm NR}}}
\newcommand{\DsPhiPi}  {\decay{\Dsm}{\phi\pim}}
\newcommand{\DsPiPiPi} {\decay{\Dsm}{\pim\pip\pim}}

\newcommand{\LbDsP}    {\texorpdfstring{\decay{\Lb}{\Dsm p}}{}}
\newcommand{\LbDsstP}  {\decay{\Lb}{\Dssm p}}
\newcommand{\LbLcK}    {\decay{\Lbbar}{\Lcbar \Kp}}
\newcommand{\LbLcPi}   {\decay{\Lbbar}{\Lcbar \pip}}
\newcommand{\LbLch}    {\decay{\Lbbar}{\Lcbar h^{+}}}

\newcommand{\hhh}      {\texorpdfstring{h^{-}h^{+}h^{\mp}}{}}

\newcommand{\mBsDsK} {\texorpdfstring{m(\DsK)}{}}

\newcommand{\mDs} {\texorpdfstring{m(\hhh)}{}}

\renewcommand{\mBsDsK} {\texorpdfstring{m(\DsK)}{}}

\renewcommand{\mDs} {\texorpdfstring{m(\hhh)}{}}

\usepackage{cite} 
\usepackage{mciteplus}

\usepackage{longtable} 
\usepackage{booktabs}
\usepackage{comment}

\begin{document}

\renewcommand{\thefootnote}{\fnsymbol{footnote}}
\setcounter{footnote}{1}


\begin{titlepage}
\pagenumbering{roman}

\vspace*{-1.5cm}
\centerline{\large EUROPEAN ORGANIZATION FOR NUCLEAR RESEARCH (CERN)}
\vspace*{1.5cm}
\noindent
\begin{tabular*}{\linewidth}{lc@{\extracolsep{\fill}}r@{\extracolsep{0pt}}}
\ifthenelse{\boolean{pdflatex}}
{\vspace*{-1.5cm}\mbox{\!\!\!\includegraphics[width=.14\textwidth]{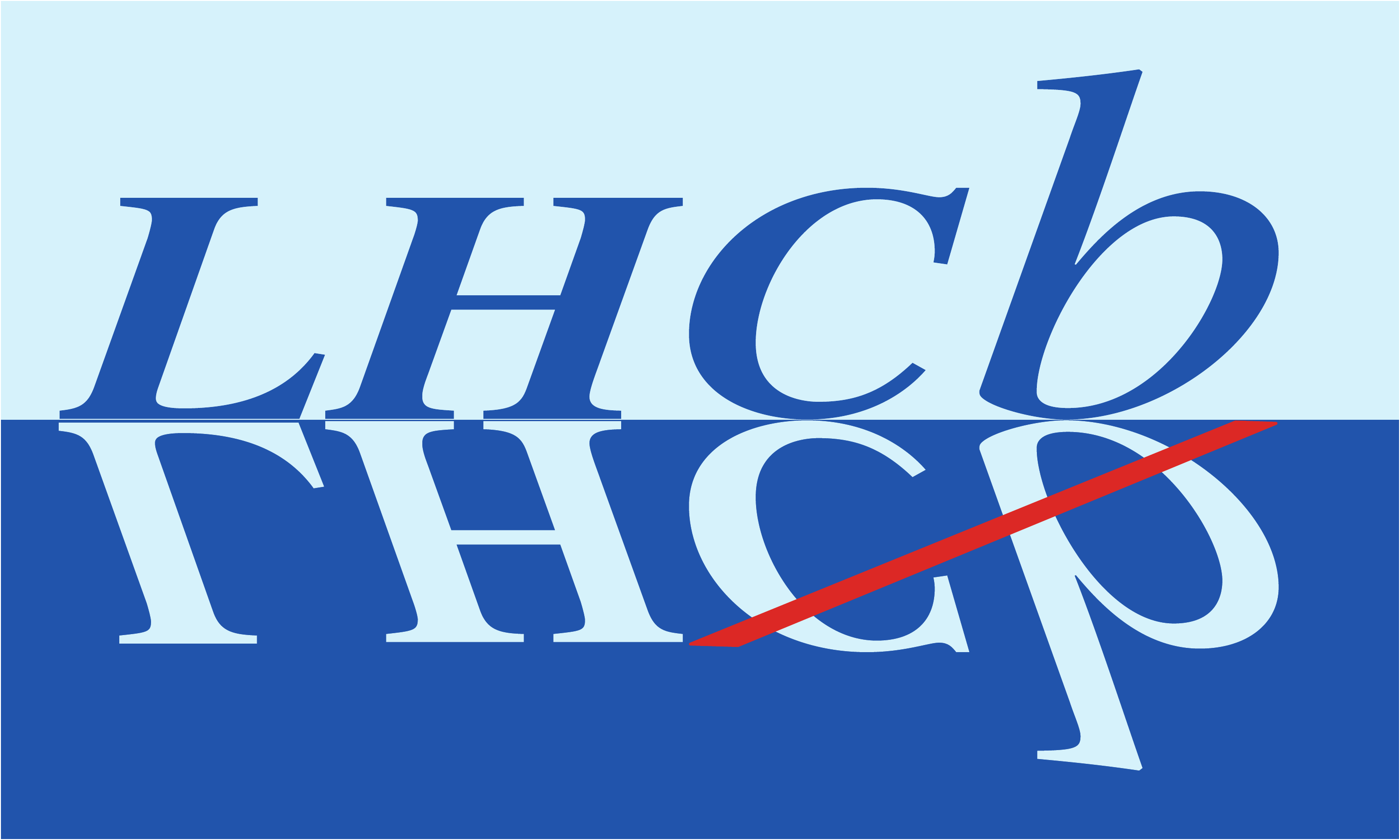}} & &}%
{\vspace*{-1.2cm}\mbox{\!\!\!\includegraphics[width=.12\textwidth]{figs/lhcb-logo.eps}} & &}%
\\
 & & CERN-EP-2024-219 \\  
 & & LHCb-PAPER-2024-020 \\  
 & & 19 March 2025 \\ 
 & & \\
\end{tabular*}

\vspace*{3.0cm}

{\normalfont\bfseries\boldmath\huge
\begin{center}
  \papertitle 
\end{center}
}

\vspace*{2.0cm}

\begin{center}
\paperauthors\footnote{Authors are listed at the end of this paper.}
\end{center}

\vspace{\fill}

\begin{abstract}
  \noindent
  A measurement of the \CP-violating parameters in \BsDsK decays is reported, based on the analysis of proton-proton collision data collected by the \lhcb experiment corresponding to an integrated luminosity
of $6 \invfb$  at a centre-of-mass energy of $13 \tev$.
The measured parameters are obtained with a decay-time dependent analysis yielding $\Cf = 0.791 \pm 0.061 \pm 0.022$, $\AdGamf = -0.051 \pm 0.134 \pm 0.058$, $\AdGamfb = -0.303 \pm 0.125 \pm 0.055$, $\Sf = -0.571 \pm 0.084 \pm 0.023$ and $\Sfb = -0.503 \pm 0.084 \pm 0.025$,
where the first uncertainty is statistical and the second systematic.
This corresponds to $\CP$ violation in the interference between mixing and decay of about $8.6\,\sigma$.
Together with the value of the \Bs mixing phase $-2\betas$, these parameters are used to obtain a measurement 
of the CKM angle $\gamma$ equal to \mbox{$ (74\pm12)^\circ$}
modulo $180^{\circ}$, where the uncertainty contains both statistical and systematic 
contributions.
This result is combined with the previous \lhcb measurement in this channel using $3 \invfb$ resulting in a determination of \mbox{$\gamma= (81^{+12}_{-11})^\circ$}.

\end{abstract}

\vspace*{1.0cm}

\begin{center}
Published in JHEP 03 (2025) 139
\end{center}

\vspace{\fill}

{\footnotesize 
\centerline{\copyright~\papercopyright. \href{\paperlicenceurl}{\paperlicence}.}}
\vspace*{2mm}

\end{titlepage}


\newpage
\setcounter{page}{2}
\mbox{~}
%
%
%
%


\renewcommand{\thefootnote}{\arabic{footnote}}
\setcounter{footnote}{0}

\cleardoublepage


\pagestyle{plain} 
\setcounter{page}{1}
\pagenumbering{arabic}


\section{Introduction}
\label{sec:Introduction}

Measurements of the \CP asymmetries in \mbox{$B^0_{(s)}\to D^{(*)\mp}_{(s)} h^\pm$}
decays,\footnote{Inclusion of charge-conjugate modes is implied 
throughout, except where explicitly stated.}
where $h=\pi$ or $K$, have been performed by the 
\lhcb~\cite{LHCb-PAPER-2017-047,LHCb-PAPER-2018-009}, \babar~\cite{Aubert:2005yf,Aubert:2006tw} 
and \belle~\cite{PhysRevD.73.092003,Bahinipati:2011yq} collaborations. These measurements
are of particular interest as they constrain elements of the CKM quark-mixing
matrix, in which all Standard Model (SM) \CP-violation effects arise from
a single complex phase~\cite{CKM1,CKM2}.
The unitarity constraint of the CKM matrix, relevant to the
$b \rightarrow u$ and $b \rightarrow c$ transitions in the above decays,
can be written as
$V_{ud}^{\phantom{*}}V_{ub}^{*} + V_{cd}^{\phantom{*}}V_{cb}^{*} + V_{td}^{\phantom{*}}V_{tb}^{*} = 0$, 
where $V_{ij}$ are the matrix elements. 
This constraint can be represented as a triangle in a complex plane in which the internal angle $\gamma$ is defined by \mbox{$\g= \phi_3 \equiv arg(-V^{\phantom{*}}_{ud}V_{ub}^{*}/V^{\phantom{*}}_{cd}V_{cb}^{*})$} and can be probed both indirectly, under the assumption of unitarity, and directly in tree-level processes~\cite{Jarlskog1985ht,JarlskogReplyCommentTOJarlskog1985ht,CommentTOJarlskog1985ht}.
The consistency between these determinations provides a powerful validation of the SM picture of \CP violation.
The most accurate determination of the angle \g in tree-level processes is currently obtained 
by combining \lhcb measurements of \Bp, \Bz and \Bs decays to final states with a $D_{(s)}$ 
meson and one or more light mesons.
Results from both time-integrated and time-dependent analyses are used, as well as constraints from charm-meson decays~\cite{LHCb-CONF-2024-004,*LHCb-PAPER-2021-033}.

\begin{figure}[tb]
  \centering 
  $\,$ \hfill \includegraphics[width=.4\textwidth]{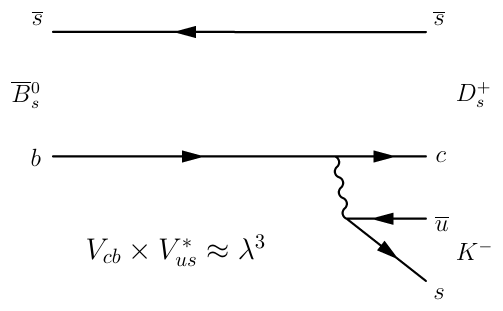} \hfill
  \includegraphics[width=.4\textwidth]{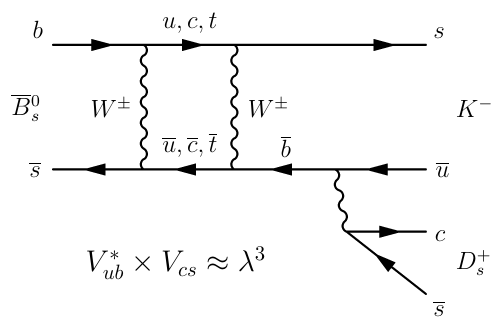} \hfill
  $\,$ \\
\vspace{-2mm}
  \caption{Leading-order Feynman diagrams for $\Bsb\to\Dsp\Km$ decays (left) without and (right) with \Bs--\Bsb mixing.}
 \label{fig:feynmandiags}
\end{figure}

The decay-time-dependent analyses of $\B^0_{s}\to \D^\mp_{s} K^{\pm}$ and $\Bd\to \D^{(*)\mp}
\pi^{\pm}$ tree-level decays
are sensitive to the angle $\g$ in the interference of mixing and decay
amplitudes~\cite{Dunietz:1987bv,Aleksan:1991nh,Fleischer:2003yb,DeBruyn:2012jp,Fleischer:2021cwb,Fleischer:2021cct}, which for the \BsDsK decay proceed through the leading-order Feynman diagrams shown in  Fig.~\ref{fig:feynmandiags}.
In these decays, the \CP-violating parameters are functions of a combination of the angle \g 
and the relevant mixing phase $\beta_{(s)}$, namely $\gamma+2\beta$ 
($\beta = \phi_1 \equiv \arg(-V^{\phantom{*}}_{cd}V_{cb}^{*}/V^{\phantom{*}}_{td}V_{tb}^{*})$)
in the \Bd system and \weak
\mbox{($\betas \equiv \arg(-V^{\phantom{*}}_{ts}V_{tb}^{*}/V^{\phantom{*}}_{cs}V_{cb}^{*})$)}
in the \Bs system.
In \mbox{$\Bd\to D^{(*)\mp}\pi^\pm$} decays the ratio between the interfering decay
amplitudes is small, \mbox{$r_{D^{(*)}\pi} = |A(\Bdb \to D^{(*)-}\pi^+)/A(\Bd \to
D^{(*)-}\pi^+)| \approx 0.02$}, which limits the sensitivity to the CKM angle 
\g~\cite{LHCb-PAPER-2020-021,PDG2024}.
By contrast, the ratio is larger for \mbox{\BsDsmKp} decays, 
\mbox{$\rdsk = |A(\Bsb \to \Dsm \Kp)/A(\Bs \to \Dsm \Kp)|$} 
$\approx 0.4$, 
since both $b\to cs\bar{u}$ and \mbox{$b\to u\bar{c}s$} amplitudes have similar magnitudes, 
of $\mathcal{O}(\lambda^3)$, where $\lambda \approx 0.23$ is the sine of the Cabibbo
angle~\cite{Wolfenstein:1983yz,PDG2024}.

This paper presents a measurement of the \CP-violating parameters in \BsDsK decays using a data set of proton-proton ($pp$) 
collisions recorded  with the \lhcb detector at a centre-of-mass energy of $\sqs = 13\tev$ 
during the Run~2 data-taking period of the LHC (2015--2018). This data set corresponds
to an integrated luminosity of $6 \invfb$.
The decays of the \Dsm meson into the final states $\Km\pip\pim$, $\pim\pip\pim$ and $\Km\Kp\pim$ are analysed.
The analysis strategy is similar to that of Ref.~\cite{LHCb-PAPER-2017-047}, with the selection, fit model and determination of systematic uncertainties reoptimised.
These improvements benefit from better trigger and reconstruction performances of 
the LHCb experiment throughout Run~2~\cite{LHCb-DP-2019-001}.

The determination of the \CP-violating parameters is achieved using a two-stage fitting procedure.
At the first stage, the \BsDsK signal is statistically separated from background components using the \emph{sPlot} technique~\cite{Pivk:2004ty}, where the signal weights are determined from a two-dimensional fit to the $\mBsDsK$ and $\mDs$ distributions. Here $h$ denotes either a kaon or a pion in the different $\Ds$ decays.
Each fit component is factorised using the product of the probability density functions (PDFs) modelling the $\mBsDsK$ and $\mDs$ invariant-mass distributions since their correlations are determined to be negligible in simulation samples. A systematic uncertainty is assigned to account for the impact of any remaining correlations.
The two-dimensional fit is performed simultaneously to all \Ds final states considered in this analysis and to three data-taking periods (2015--2016, 2017, 2018), where the 2015 sample is fitted together with 2016 data due to its limited size.  
In the second stage, an unbinned maximum-likelihood fit to the decay-time distribution of the background-subtracted \BsDsK signal is performed to determine the \CP-violating parameters. In the decay-time fit the data-taking periods are fitted simultaneously, while the \Ds final states are combined. 

Finally, the results of the present analysis are combined with those of 
Ref.~\cite{LHCb-PAPER-2017-047}, which uses an integrated luminosity of $3 \invfb$ recorded 
at $\sqs = 7$ and $8 \tev$ during the Run~1 data-taking period (2011--2012) with the external inputs updated 
to match the values used in the present analysis.

\vspace{-2mm}

\subsection{Decay rates and \texorpdfstring{$\boldsymbol{\CP}$}{CP}-violating parameters}

\label{sec:equations}
Following the conventions of Ref.~\cite{LHCb-PAPER-2017-047},
the time-dependent decay rates of an initially produced flavour eigenstate
$\Bs$ or $\Bsb$ decaying to final state $f$ can be written as
\begin{align}
\frac{{\rm d}\Gamma_{\Bs\to\f}(t)}{{\rm d}t} &= \frac{1}{2} |\Af|^2 (1+|\lf|^2) e^{-\Gs t} \left[
         \cosh\left(\frac{\DGs t}{2}\right) 
  + \AdGamf\sinh\left(\frac{\DGs t}{2}\right) \right. \nonumber\\
& + \Cf\cos\left(\dms t\right) 
  - \Sf\sin\left(\dms t\right)
\Big] \ ,
\label{eq:decay_rates_1}\\
\frac{{\rm d}\Gamma_{\Bsb\to\f}(t)}{{\rm d}t} &= \frac{1}{2} |\Af|^2 \left|\frac{p}{q}\right|^2 (1+|\lf|^2) e^{-\Gs t} \left[
         \cosh\left(\frac{\DGs t}{2}\right) 
  + \AdGamf\sinh\left(\frac{\DGs t}{2}\right) \right. \nonumber\\
& - \Cf\cos\left(\dms t\right) 
  + \Sf\sin\left(\dms t\right)
\Big] \ ,
\label{eq:decay_rates_2}
\end{align}
where \mbox{$\lf \equiv (q/p)(\Abf/\Af)$}, $\Af (\Abf)$ is the amplitude of a \Bs (\Bsb) 
decay to the final state $f \equiv D_s^- K^+$  and the complex coefficients $p$ and $q$
describe the mixing of the light, $|B_{L}\rangle$, and heavy, $|B_{H}\rangle$, mass 
and flavour eigenstates according to
\begin{equation}
\begin{aligned}
|B_L\rangle \equiv p|\Bs\rangle+q|\Bsb\rangle \;\;\; {\text{and} }\;\;\; 
|B_H\rangle \equiv p|\Bs\rangle-q|\Bsb\rangle \ ,
\label{eq:mixing}
\end{aligned}
\end{equation} with the normalisation condition $|p|^2+|q|^2 = 1$.
Here, \Gs is the  \Bs decay width or inverse \Bs lifetime, $\DGs \equiv\Gamma_{B_L} - \Gamma_{B_H}$
is the decay-width difference between the light and heavy mass eigenstates and 
$\dms \equiv m_{B_{H}} - m_{B_{L}}$ is the mixing frequency in the \Bs system.
Similar equations can be obtained
for decays into the \CP-conjugate final state $\bar{f} \equiv D_s^+ K^-$ by replacing \Cf by 
\Cfb, \Sf by \Sfb, and \AdGamf by \AdGamfb.
The \CP-asymmetry parameters can be written as
\begin{equation}
\begin{aligned}
\Cf  &=  \frac{ 1 - |\lf|^2 }{ 1 + |\lf|^2 }  = \hspace{+0.1cm} -\Cfb = - \frac{1-|\lfb|^2}{1+|\lfb|^2}\ , \\
\AdGamf  &= \frac{ -2 \Real(\lf) }  { 1 + |\lf|^2 } \ , \quad
\hspace{-0.3cm}\AdGamfb = \frac{ -2 \Real(\lfb) }{ 1 + |\lfb|^2 } \ , \\
\Sf  &= \frac{ 2 \Imag(\lf) }  { 1 + |\lf|^2 } \ ,\quad
\hspace{+0.25cm} \Sfb = \frac{ 2 \Imag(\lfb) }{ 1 + |\lfb|^2 } \ .
\label{eq:asymm_obs}
\end{aligned}
\end{equation}
The relation $\Cf = -\Cfb$ results from the conditions $|q/p| = 1$ and $|\lf| = |1/\lfb|$, which imply no \CP violation both in mixing, in agreement with current measurements 
\cite{LHCb-PAPER-2016-013}, and decay. The second assumption is motivated by the fact that
only a single amplitude contributes to each initial-to-final-state transition. Finally, 
the \CP observables are related to the magnitude of the amplitude ratio
$\rdsk$,
the strong-phase difference \strong between the amplitudes $A(\Bsb \to \Dsm \Kp)$ and
\mbox{$A(\Bs \to \Dsm \Kp)$} and the weak-phase difference \weak by the following equations
\begin{equation}
\begin{aligned}
\Cf  	 = &\frac{1-\rdsk^2}{1+\rdsk^2}                   \ ,   \\
\AdGamf  = \frac{-2 \rdsk \cos(\strong-(\weak))}{1+\rdsk^2}\ &, \quad
\AdGamfb = \frac{-2 \rdsk \cos(\strong+(\weak))}{1+\rdsk^2}\ ,  \\
\Sf 	 = \frac{2 \rdsk \sin(\strong-(\weak))}{1+\rdsk^2}\ &, \quad
\Sfb	 = \frac{-2 \rdsk \sin(\strong+(\weak))}{1+\rdsk^2}\ .
\label{eq:truth}
\end{aligned}
\end{equation}
These observables are used to extract $\gamma$, $\strong$ and $\rdsk$ while fixing $-2\beta_s$, as discussed in Sec.~\ref{sec:interpretation}.
The combined Run 1 and Run 2 result is also expressed in terms of $\weak$. 
This combined quantity offers complementary sensitivity on a potential new physics phase in $\Bs$--$\Bsb$ mixing.

\section{Detector and software}
\label{sec:Detector}

The \lhcb detector~\cite{LHCb-DP-2008-001,LHCb-DP-2014-002} is a single-arm forward
spectrometer covering the \mbox{pseudorapidity} range $2<\eta <5$,
designed for the study of particles containing \bquark or \cquark
quarks. The detector includes a high-precision tracking system
consisting of a silicon-strip vertex detector surrounding the $pp$
interaction region, a large-area silicon-strip detector located
upstream of a dipole magnet with a bending power of about
$4{\mathrm{\,T\,m}}$, and three stations of silicon-strip detectors and straw
drift tubes placed downstream of the magnet.
The tracking system provides a measurement of the momentum, \ptot, of charged particles with
a relative uncertainty that varies from 0.5\% at low momentum to 1.0\% at 200\gevc.
The minimum distance of a track to a primary $pp$ collision vertex (PV), the impact parameter (IP), 
is measured with a resolution of $(15+29/\pt)\mum$,
where \pt is the component of the momentum transverse to the beam, in\,\gevc.
Different types of charged hadrons are distinguished using information
from two ring-imaging Cherenkov detectors. 
Photons, electrons and hadrons are identified by a calorimeter system consisting of
scintillating-pad and preshower detectors, an electromagnetic
and a hadronic calorimeter. Muons are identified by a
system composed of alternating layers of iron and multiwire
proportional chambers.
The online event selection is performed by a trigger, 
which consists of a hardware stage, based on information from the calorimeter and muon
systems, followed by a software stage, which applies a full event
reconstruction.

At the hardware trigger stage, events are required to have a muon with high \pt or a
hadron, photon or electron with high transverse energy in the calorimeters.
The software trigger requires a two-, three- or four-track
secondary vertex with a significant displacement from any primary
$pp$ interaction vertex. At least one charged particle
must have a transverse momentum $\pt > 1.6\gevc$ and be
inconsistent with originating from a PV.
A multivariate algorithm~\cite{BBDT,LHCb-PROC-2015-018} is used for
the identification of secondary vertices consistent with the decay
of a \bquark hadron.

Simulation is required to model the effects of the detector acceptance and the
imposed selection requirements.
In the simulation, $pp$ collisions are generated using
\pythia~\cite{Sjostrand:2007gs,*Sjostrand:2006za} 
with a specific \lhcb configuration~\cite{LHCb-PROC-2010-056}.
Decays of unstable particles
are described by \evtgen~\cite{Lange:2001uf}, in which final-state
radiation is generated using \photos~\cite{davidson2015photos}.
The interaction of the generated particles with the detector, and its response,
are implemented using the \geant
toolkit~\cite{Allison:2006ve, *Agostinelli:2002hh} as described in
Ref.~\cite{LHCb-PROC-2011-006}.

\section{Candidate selection}
\label{sec:selection}

The selection criteria are similar to those used in Ref.~\cite{LHCb-PAPER-2017-047}, but
updated to reflect improvements in reconstruction performance.
Samples of signal  \BsDsK and two control channels \BsDsPi and \BdDPi decays are selected by combining 
a $D_{(s)}^-$ candidate with a particle, referred to as ``companion'' in the following, 
consistent with the hypothesis of being either a kaon or a pion.
The \Dm meson is reconstructed using the \DKPiPi decay.
The \Dsm meson is reconstructed using the final states $\Km\pip\pim$, $\pim\pip\pim$ and $\Km\Kp\pim$, with the latter further subdivided into \DsPhiPi, $\Dsm \to K^{*}(892)^{0} \Km$, and the remaining regions of the phase space denoted as the nonresonant \DsNonRes component.
The separation between these five decay modes is based on kinematic and particle identification (PID) requirements, allowing for the optimisation of the signal selection while accounting for the different background contributions in each sample.
Furthermore, each decay mode undergoes a combination of PID and kinematic vetoes to suppress cross-feed backgrounds from \BdDh or \LbLch decays, as well as background contributions containing $\jpsi$, $\Dz$ and $\Kstarz$ decays.
The distinction between \BsDsPi and \BsDsK decays is achieved with mutually exclusive requirements on the PID information of the companion track.

The \Bs candidate is associated with the PV with the smallest impact parameter $\chi^2$, calculated as the difference in $\chi^2$ for the vertex fit of the PV with and without the considered particle (\chisqip).
The decay-time resolution of the \Bs candidate is enhanced through a kinematic fit~\cite{Hulsbergen:2005pu}, which constrains the candidate to originate from the associated PV. Similarly, the measured values of the $\DsK$ invariant mass are obtained by constraining the $\Dsm$ invariant mass to the world-average value~\cite{PDG2024}. 
The \Bs and \Dsm candidates are required to have invariant masses within $[5300,5800]\mevcc$ and $[1930,2015]\mevcc$, respectively.

Contributions from $b$-hadron decays that do not include a charm hadron are suppressed by imposing a \Dsm flight-distance significance requirement. This quantity is defined as the distance between the \Bs and \Dsm decay vertices divided by its uncertainty.

To suppress combinatorial background due to random track combinations, a gradient-boosted decision tree (BDTG) algorithm~\cite{Breiman,Roe} is employed.
The BDTG classifier is trained using the \mbox{\BsDsPi} control sample reconstructed in the $\DsKKPi$ final state, as detailed in Ref.~\cite{LHCb-PAPER-2021-005}. Since all channels in this analysis exhibit similar kinematics and no PID information is used in the BDTG classifier, the resulting BDTG algorithm performs equally well on other \Dsm decay modes.
The classifier uses several track-related variables, including the transverse momentum of the companion particle, the radial flight distance of both the $b$- and $c$-hadron candidates and the companion and $b$-hadron's minimum $\chisqip$. A detailed description can be found elsewhere~\cite{LHCb-PAPER-2017-047,HeinickeKevin}.
Fewer than 0.5\% of the events passing the selection requirements contain more than one signal candidate, and in such cases, all candidates are used in the analysis.

\section{Two-dimensional invariant-mass fit}
\label{sec:mdfit}
The selected $\BsDsK$ candidates are fitted using a two-dimensional unbinned extended maximum-likelihood fit to the $\mBsDsK$ and $\mDs$ distributions, in order to statistically remove background components using the \emph{sPlot} technique in the subsequent decay-time fit.

The signal and background PDFs for the invariant-mass fit are derived from the simulated samples after being corrected to better reproduce data.
Specifically, the \BdDPi control mode is used to correct for differences between simulation and data in the distributions of the $\Bs$ and $\Ds$ vertices' uncertainty on the $z$-position and to correct for a shift between data and simulation in the $\mBsDsK$ invariant mass.
The PID distributions in the simulation are corrected to match those in data using $\Dstarp\to\Dz\pip$ and $\Lz^0\to \proton\pim$ calibration samples. More information about this procedure is provided in Ref.~\cite{LHCb-DP-2018-001}.

The shape of the $\mBsDsK$ distribution for signal candidates is modelled using the sum of a double-sided Hypatia function~\cite{Santos:2013gra} and a Johnson $S_U$ function~\cite{johnson1949systems}, sharing a common peak position. This combination effectively describes the main peak and the radiative tail.
The signal PDFs are separately derived from simulated \BsDsK candidates for each \Dsm decay mode and data-taking period.
The signal shapes are fixed in the data fit with two exceptions. Separate peak parameters are used for the three data-taking periods, which are left free in the fit. 
Furthermore, the widths of the Hypatia functions are fixed to the values determined from simulation, while the widths of the Johnson $S_U$ functions are left free.
This adjustment compensates for mass-resolution differences between simulation and data.

The $\mDs$ signal distribution is also described using the combination of a double-sided Hypatia function and a Johnson $S_U$ function, sharing a common peak position.
The signal PDFs are derived from simulation for each \Dsm decay mode and data-taking period.
Similar to the $\mBsDsK$ invariant-mass parameterisation, only the common peak position and the width of the Johnson $S_U$ function are free parameters in the fit to data.

The combinatorial background comprises random track combinations that do not originate from a \Dsm meson decay, as well as backgrounds containing a true \Dsm decay combined with a random companion track.
The functional form of the combinatorial background is motivated by the upper $\mBsDsK$ invariant-mass sidebands, $[5600,6800]$\mevcc, with all parameters allowed to vary in the two-dimensional fit.
It is parameterised separately for each \Dsm decay mode and each period of data taking: an exponential plus constant is found to be optimal to describe the combinatorial background component for the \DsKKPi decay modes, while for the \DsKPiPi and \DsPiPiPi final states a single exponential is found to be sufficient.
The combinatorial component in the $\mDs$ distribution is described by the sum of an exponential and the \Ds signal shape, where the peak position is shared with the signal itself, modelling the contributions from random track combinations and true $\Ds$ mesons paired with a random track, respectively.
The exponent and the fraction between the exponential and the signal shape are free to vary in the fit to data.

In the two-dimensional fit, besides the signal and the combinatorial background, the following background contributions are considered: fully reconstructed \BdDsK decays, companion-track misidentified $\BsDsPi$ and $\LbDsP$ decays, companion-track misidentified partially reconstructed $\BsDsstPi$, $\BsDsRho$ and $\LbDsstP$ decays, where the neutral $\gamma$ or $\piz$ from $\Dssm\to \Dsm \gamma/\piz$ and $\rhop\to\pip\piz$ is not reconstructed.
Furthermore, the $\BdDK$, $\BdDPi$, $\LbLcK$ and $\LbLcPi$ components are included, where a misidentified final state causes the $c$-hadron to be reconstructed as a $\Dsm$ meson.

In the fit to the $\mBsDsK$ distribution, double-sided Hypatia functions are used for the fully reconstructed \mbox{\BdDsK} and the partially reconstructed \BsDsstPi contributions,
the sum of a double-sided Hypatia and a Johnson $S_U$ function is used to describe the \BsDsPi decays,
the contribution from partially reconstructed \BsDsRho decays is modelled using the sum of two exponential functions. 
For the $\mBsDsK$ distribution of the remaining partially or fully reconstructed backgrounds, the shapes of the distributions in the simulation are defined using a nonparametric kernel estimation method~\cite{Cranmer:2000du}, corrected to match the PID efficiency and kinematic distributions observed in the data.
In the fit to the $\mDs$ distribution, the signal shape is used for the \BdDsK, \BsDsPi, \BsDsstPi, \BsDsRho, \LbDsP and \LbDsstP contributions, whereas the other background components are described using a nonparametric kernel estimation method.

\begin{figure}[!tb]
  \centering
  \includegraphics[width=\textwidth]{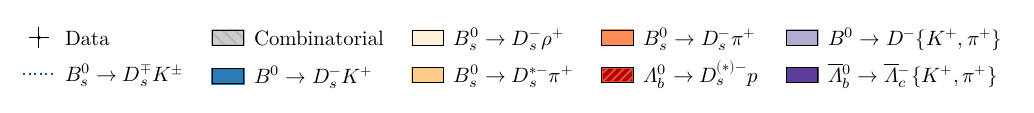}
  \includegraphics[width=.485\textwidth]{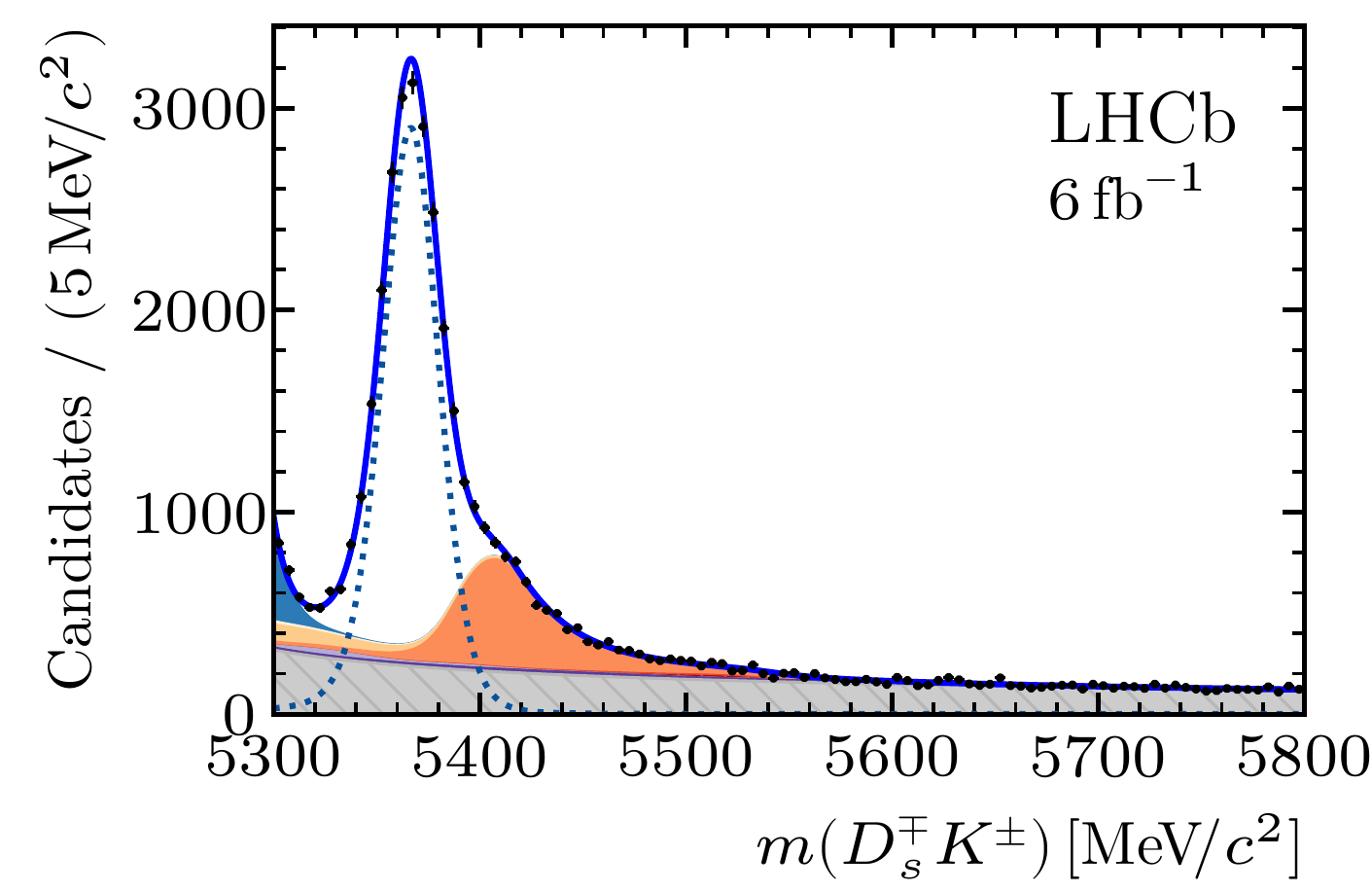}
  \includegraphics[width=.485\textwidth]{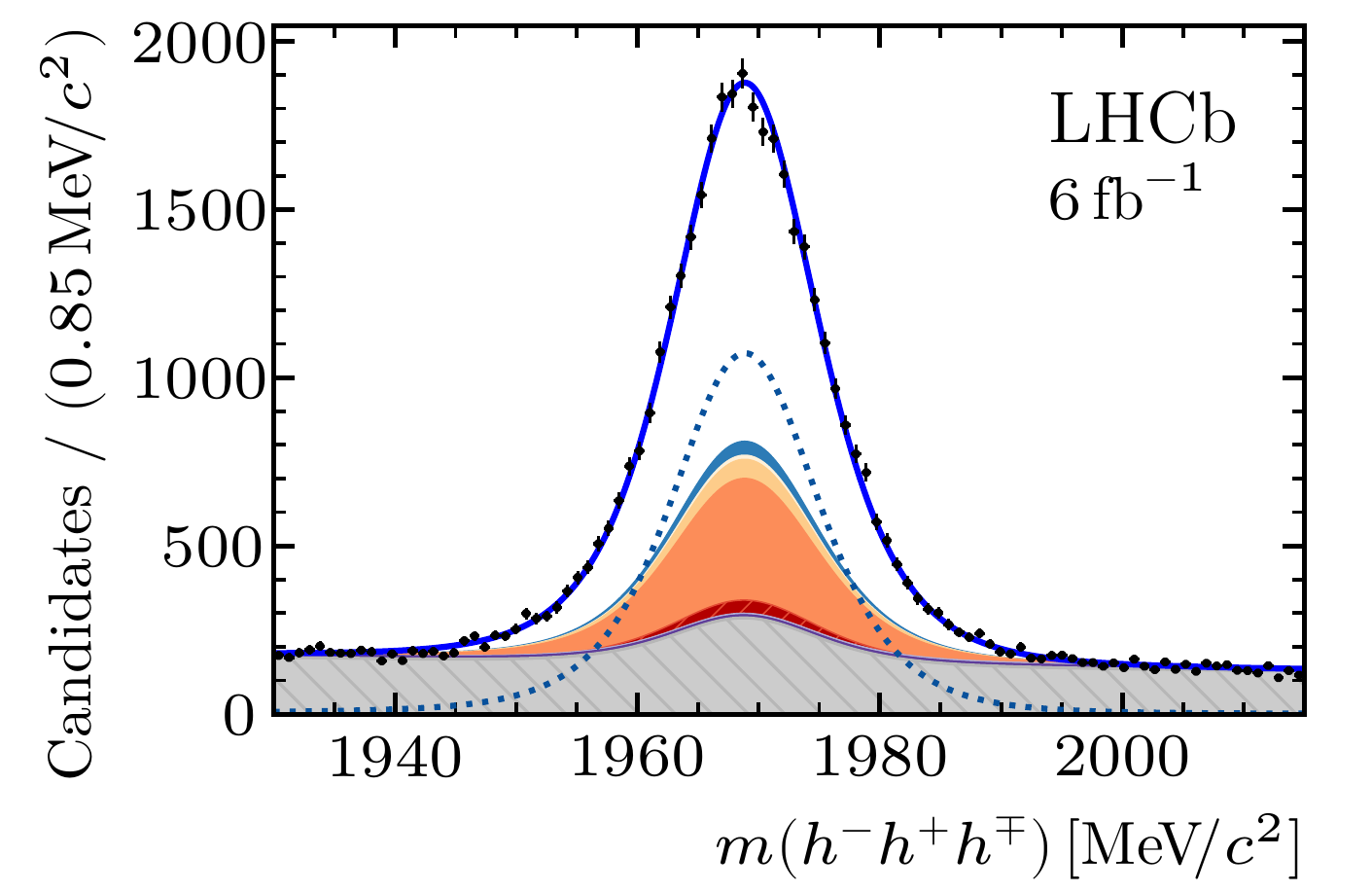}
  \caption{Distributions of the (left) $\Dsmp K^{\pm}$ and (right) $\hhh$ invariant masses for selected $\DsK$ candidates.
  The contributions from all \Dsm decay modes and years of data taking are combined. The solid-blue curve is
  the total result of the two-dimensional fit. The dotted curve shows the \BsDsK signal and the
  shaded-stacked histograms show the different background contributions.}
  \label{fig:massfit-BsDsK-All}
\end{figure}

The invariant-mass fit is performed simultaneously across the different \Dsm decay modes and periods of data taking.
For each \Dsm decay mode, the PDF is constructed from the sum of signal and background contributions.
Most background yields are allowed to vary freely in the fit, except for those with an expected contribution below 2\% of the signal yield, specifically: \BdDK, \BdDPi, \LbLcK and \LbLcPi. 
In such cases, the yields are fixed to the results of dedicated fits to $\BdDPi$ and $\LbLcPi$ candidates, corrected by the known branching fractions and selection efficiencies relative to the selection of the $\BsDsK$ candidates~\cite{PDG2024}.

The $\Dsmp K^{\pm}$ and $h^\mp h^+ h^-$ invariant-mass distributions, summed over the \Dsm decay modes and the data-taking periods, are shown in Fig.~\ref{fig:massfit-BsDsK-All} with the results of the fit overlaid.
The signal yield, confirmed to be unbiased through data-like pseudoexperiments, is determined to be $20\,949 \pm 180$, where the uncertainty is statistical only.
The invariant-mass fit to the \BsDsPi control mode is reported in Ref.~\cite{LHCb-PAPER-2021-005}. 

\section{Decay-time fit}
\label{sec:timefit}

The goal of the decay-time fit is to determine the $\CP$-violating parameters $\Cf$, $\Sf$, $\Sfb$, $\AdGamf$ and $\AdGamfb$ from the decay-time distribution of $\BsDsK$ decays which follow a PDF obtained from Eqs.~\ref{eq:decay_rates_1}~and~\ref{eq:decay_rates_2}.
The decay-time distribution of the signal candidates in the reconstructed $\BsDsK$ decays is obtained using the {\emph{sPlot}} technique following the two-dimensional mass fit.
Several experimental effects are accounted for in the fit to describe the observed decay rates, such as the decay-time resolution and acceptance and the performance of the flavour-tagging algorithms to determine the initial \Bs flavour.

To account for the decay-time resolution, the signal PDF is convolved with a Gaussian function whose width, $\sigma_t$, is evaluated for each candidate based on the decay-time uncertainty estimated by the vertex fit, $\delta_t$.
The decay-time resolution model is calibrated using a control sample of artificial $\BsDsK$ decays obtained by combining a $\Dsm$ meson originating promptly from the $pp$ interaction point with an oppositely charged companion meson coming from the same origin. 
The selection requirements are the same as for the signal, except for those that depend on the displacement from the $pp$ collision point.
These $\Bs$ candidates have a decay vertex compatible with a lifetime of zero, up to resolution and alignment effects, thus their decay-time distribution determines the decay-time resolution.
The sample of prompt $\Ds$ candidates is divided into ten equally populated bins of decay-time uncertainty.
In each subsample, the decay-time resolution is determined from a fit to the decay-time distribution.
The resulting decay-time resolutions as a function of decay-time uncertainty are then fitted with a linear function, $\sigma_t = p_0 + p_1 \cdot \delta_t$.
The resulting average decay-time resolution is about $46\fs$.

A bias in the measured \Bs decay time, originating from residual detector misalignment, has been identified in the analysis of $\BsDsPi$ decays for the \dms measurement~\cite{LHCb-PAPER-2021-005}.
This bias was determined from the analysis of the same calibration sample of prompt $\Dsm$ mesons used for the study of the aforementioned decay-time resolution model.
Corrections for the differences between the prompt $\Dsm$ and signal samples were obtained using intentionally misaligned simulated samples.
This measurement takes the bias obtained in the $\dms$ measurement with a correction to account for differences in selection criteria between the two analyses.
The decay-time bias, determined for each period of data taking, is corrected for in the decay-time fit and amounts to about $-3\fs$.

The initial flavour of the \Bs meson is needed to determine its contribution to the corresponding decay rate. 
This information is provided by flavour-tagging algorithms which exploit different processes correlated with the \bquark-hadron production in $pp$ collisions.
Beauty quarks are predominantly produced through \bquark\bquarkbar pairs.
While one of these \bquark quarks leads to the signal \Bs meson, the other leads to a \bquark hadron that decays independently.
The decay chain of the other \bquark hadron is exploited by the opposite-side (OS) tagging algorithms~\cite{LHCb-PAPER-2011-027} to determine the initial flavour of the signal \Bs meson.
The OS muon and OS electron taggers exploit the semileptonic decay of the \bquark hadron, and the OS kaon and the OS charm taggers identify remnants from $b \rightarrow c \rightarrow s$ and $b \rightarrow c$ transitions, respectively.
Furthermore, the OS vertex-charge tagger reconstructs an effective charge of a displaced vertex from the OS \bquark-hadron decay~\cite{LHCb-PAPER-2015-027}.
Each of these algorithms infers the initial \Bs meson flavour from the charge of either a reconstructed tagging particle or, in the case of the OS vertex tagger, of a reconstructed vertex.
Additionally, the same-side (SS) kaon tagger determines the initial flavour of the \Bs signal from the charge of kaons originating from the $\squark\squarkbar$ pair produced in the fragmentation process that leads to the signal \Bs meson~\cite{LHCb-PAPER-2015-056}.
In addition to the tag decision, representing the determined flavour, the algorithms provide an estimate of the probability that the decision is wrong, the estimated mistag probability, $\eta$.
This estimate does not necessarily match the correct mistag probability, $\omega$, of the data sample.
Hence, a calibration is performed using a control sample of flavour-specific \BsDsPi decays to provide the measured mistag following the procedure described in Ref.~\cite{LHCb-PAPER-2021-005}.
The calibration functions are defined as
\begin{align}
\label{eq:mistag-1}
        \omega^{{\rm tag},y}(\eta^{{\rm tag},y}) &= \sum_{i=0}^1
        \left(f_i^{{\rm tag},y} + \frac 1 2 \Delta f_i^{{\rm tag},y} \right)
        \cdot \left(\eta^{{\rm tag},y} - \langle\eta\rangle^{{\rm tag},y}\right)^i\qquad\text{for }\Bs \ , \\
        \label{eq:mistag-2}
        \text{and }\qquad\overline{\omega}^{{\rm tag},y}(\eta^{{\rm tag},y}) &= \sum_{i=0}^1
        \left(f_i^{{\rm tag},y} - \frac 1 2 \Delta f_i^{{\rm tag},y} \right)
        \cdot \left(\eta^{{\rm tag},y} - \langle\eta\rangle^{{\rm tag},y}\right)^i\qquad\text{for }\Bsb \ ,
\end{align}
where $\langle\eta\rangle^{{\rm tag},y}$ is the average mistag probability, index $i$ identifies each of the
two flavour-tagging calibration parameters $f_i^{{\rm tag},y}$.
The parameters $\Delta f_i^{{\rm tag},y}$ are introduced to allow for different calibrations for \Bs and \Bsb candidates.
Two taggers, the SS kaon tagger ($\rm{tag} = \rm{SS}$) and the OS combination ($\rm{tag} = \rm{OS}$),  are independently calibrated on each data sample $y$ (2015--2016, 2017, 2018).
Here, the OS combination is a single set of tag decisions and mistag estimates obtained from the combination of all individual OS algorithms following the approach described in Ref.~\cite{ LHCb-PAPER-2011-027}.

The tagging information is included in the PDF used in the decay-time fit, for which
the tagging decision assigns candidates to the corresponding decay rate described by Eqs.~\ref{eq:decay_rates_1}~and~\ref{eq:decay_rates_2}.
The mistag probability causes a reduction of the oscillation amplitude by a dilution factor $D=(1-2\omega)$.
In the fit, the calibration is performed by constraining the function of the predicted mistag described by Eqs.~\ref{eq:mistag-1}~and~\ref{eq:mistag-2}.

The tagging efficiency of the full sample is $\varepsilon=(80.30\pm0.07)\%$ with an average mistag fraction of $\omega=(36.21\pm0.02\pm0.17)\%$, where the first uncertainty is due to the finite size of the calibration sample and the second is due to the uncertainty of the calibration parameters.
This results in a tagging power of $(6.10\pm0.02\pm0.15)\%$, which indicates the remaining statistical power of the flavour-tagged sample, relative to a perfectly tagged sample.
For comparison, the tagging power achieved in Run~1 was $(5.80\pm0.25)\%$~\cite{LHCb-PAPER-2017-047}.

Since the \CP-violating parameters depend on the decay-time acceptance, the latter needs to be determined.
For flavour-specific \BsDsPi decays, where $\Cf=-\Cfb = 1$ and $\Sf = \Sfb =0$, the decay-time acceptance can be determined from a fit to the decay-time distribution with \Gs and \DGs parameters fixed to the combination of LHCb results\cite{LHCb-PAPER-2023-016}.
In the \BsDsK fit, the decay-time acceptance is fixed to the result obtained from the \BsDsPi data fit, corrected by the decay-time acceptance ratio of the two modes estimated from simulation, which is weighted to match the data as described in Sec.~\ref{sec:mdfit}. 
The decay-time acceptance is modelled using segments of cubic B-splines, which are implemented analytically in the decay-time fit~\cite{Karbach:ComplxErrFunc}.
The spline boundaries, also known as knots, are chosen in order to accurately model the features of the decay-time acceptance shape. The signal decay-time PDF is then adjusted by multiplying by the decay-time acceptance model.

The decay-time fit requires additional inputs in the form of the following parameters
\begin{equation}
\begin{aligned}
\Delta m_s &= (17.7683 \pm 0.0057)\invps \ , \\ 
\Gamma_s &= (0.6563 \pm 0.0020)\invps \ ,  \\
\Delta\Gamma_s &= (0.085 \pm 0.004)\invps \ ,  \\
A_{\textrm{prod}} &= (-0.33 \pm 0.32)\% \ ,   \\
A_{\textrm{det}} &= (0.96 \pm 0.15)\% , 
\end{aligned}
\label{eq:decay-fit-inputs}
\end{equation}
which are fixed to their central values in the baseline decay-time fit and varied within their uncertainties to determine the associated systematic uncertainties. 
The values of $\Delta m_s$, $\Gamma_s$ and $\Delta \Gamma_s$ are based on LHCb measurements~\cite{LHCb-PAPER-2021-005,LHCb-PAPER-2023-016}.
The production asymmetry, $A_{\textrm{prod}}$, is fixed to the value obtained in Ref.~\cite{LHCb-PAPER-2021-005} and is defined as the relative difference in the $\Bs$ and $\Bsb$ production cross-section.
The parameter $A_{\textrm{det}}$ is defined as the relative difference in detection efficiencies between the $\Dsm K^+$ and the $\Dsp K^-$ final states.
This is evaluated following the method described in Ref.~\cite{LHCb-PUB-2018-004}, where the detection asymmetry is evaluated for $K^+\pim$ pairs using $\Dp\to \Km\pip\pip$ and $\Dp\to \Kzb \pip$ decays.
For this measurement, an average asymmetry over the \Dsm final states is calculated using the signal yields obtained from the invariant-mass fit.
Its value is compatible with the $K^+\pim$-pair detection asymmetry in Ref.~\cite{LHCB-PAPER-2020-036}, where the asymmetry was evaluated for $\Bp \to D \pip$ decays.
The detection and the production asymmetries contribute to the decay-time PDF with multiplicative factors of $(1 \pm A_{\textrm{prod}})$ and $(1 \pm A_{\textrm{det}})$ to the decay rates defined by Eqs.~\ref{eq:decay_rates_1}~and~\ref{eq:decay_rates_2}, depending on the tagged initial state and the reconstructed final state.

The \CP-violating parameters are determined in a weighted maximum-likelihood fit to the flavour-tagged decay-time distributions.
The fit is performed simultaneously to all five \Dsm final states and three data-taking periods, where calibrations of the decay-time resolution and the mistag probability are constrained individually for each data-taking period.
The fitted covariance matrix is corrected following an asymptotically correct approach described in Ref.~\cite{Langenbruch_2022} to provide good coverage of the uncertainties following the use of the \emph{sPlot} method.
The resulting \CP-violating parameters are listed in Table~\ref{tab:timefit_bsdsk} and
the corresponding statistical correlation matrix is given in
Table~\ref{tab:timefit_bsdskcorr}.

\begin{table}[!tb]
\centering
\caption{Values of the \CP-violating parameters obtained from the decay-time fit to \BsDsK candidates.
The first uncertainty is statistical and the second is systematic. }
\label{tab:timefit_bsdsk}
\begin{tabular}{lc}
  \toprule
  Parameter     & Value \\
  \midrule
  $\quad$ \Cf         & $\phantom{+}0.791  \pm  0.061 \pm 0.022$  \\ 
  $\quad$ \AdGamf     & $-0.051  \pm  0.134 \pm 0.058$  \\
  $\quad$ \AdGamfb    & $-0.303  \pm  0.125 \pm 0.055$  \\
  $\quad$ \Sf         & $-0.571  \pm 0.084 \pm 0.023$  \\
  $\quad$ \Sfb        & $-0.503  \pm 0.084 \pm 0.025$  \\
  \bottomrule
\end{tabular}
\end{table}

\begin{figure}[!tb]
 \centering
 \includegraphics[width=0.66\textwidth]{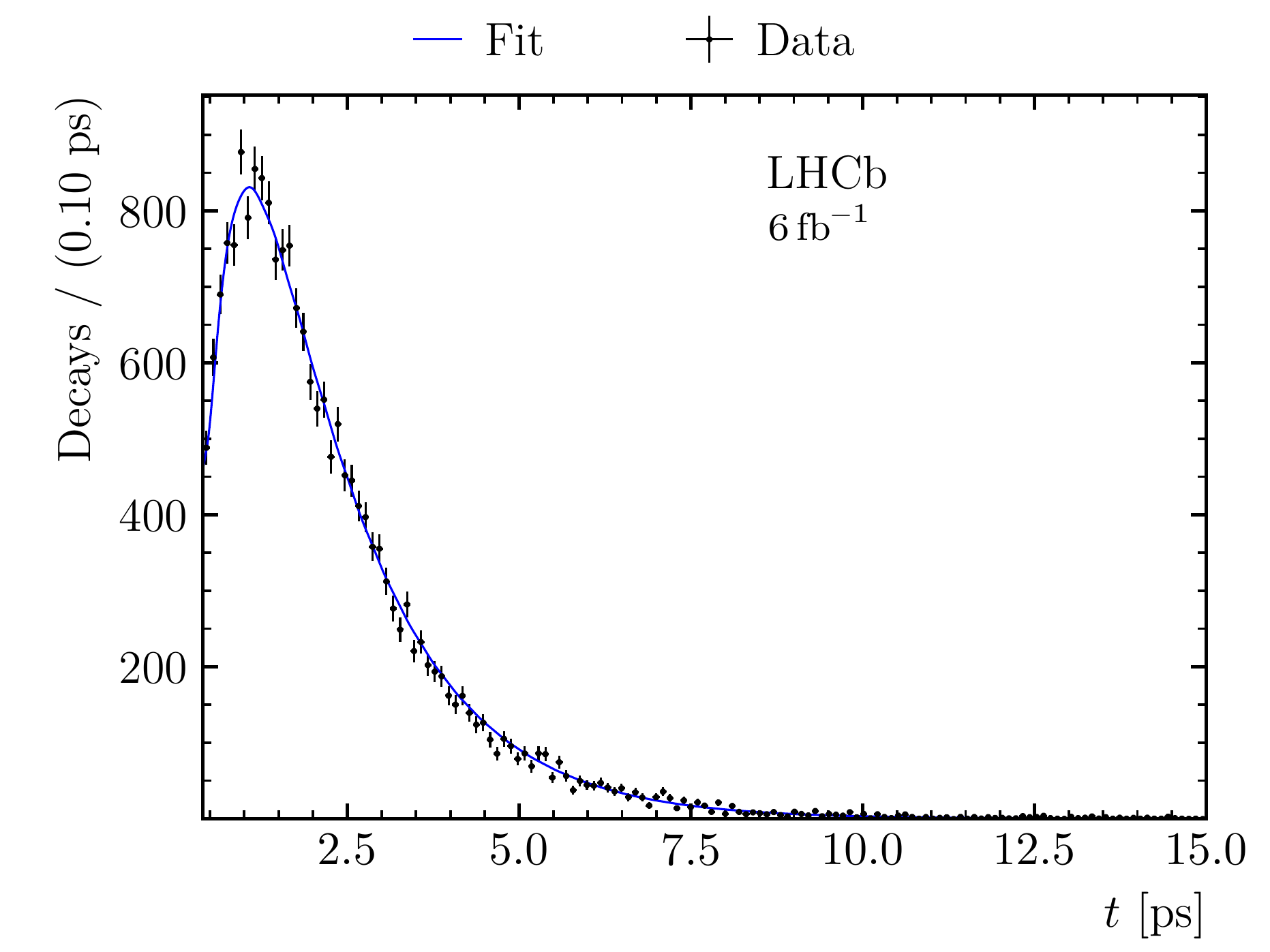} \\
 \includegraphics[width=0.66\textwidth]{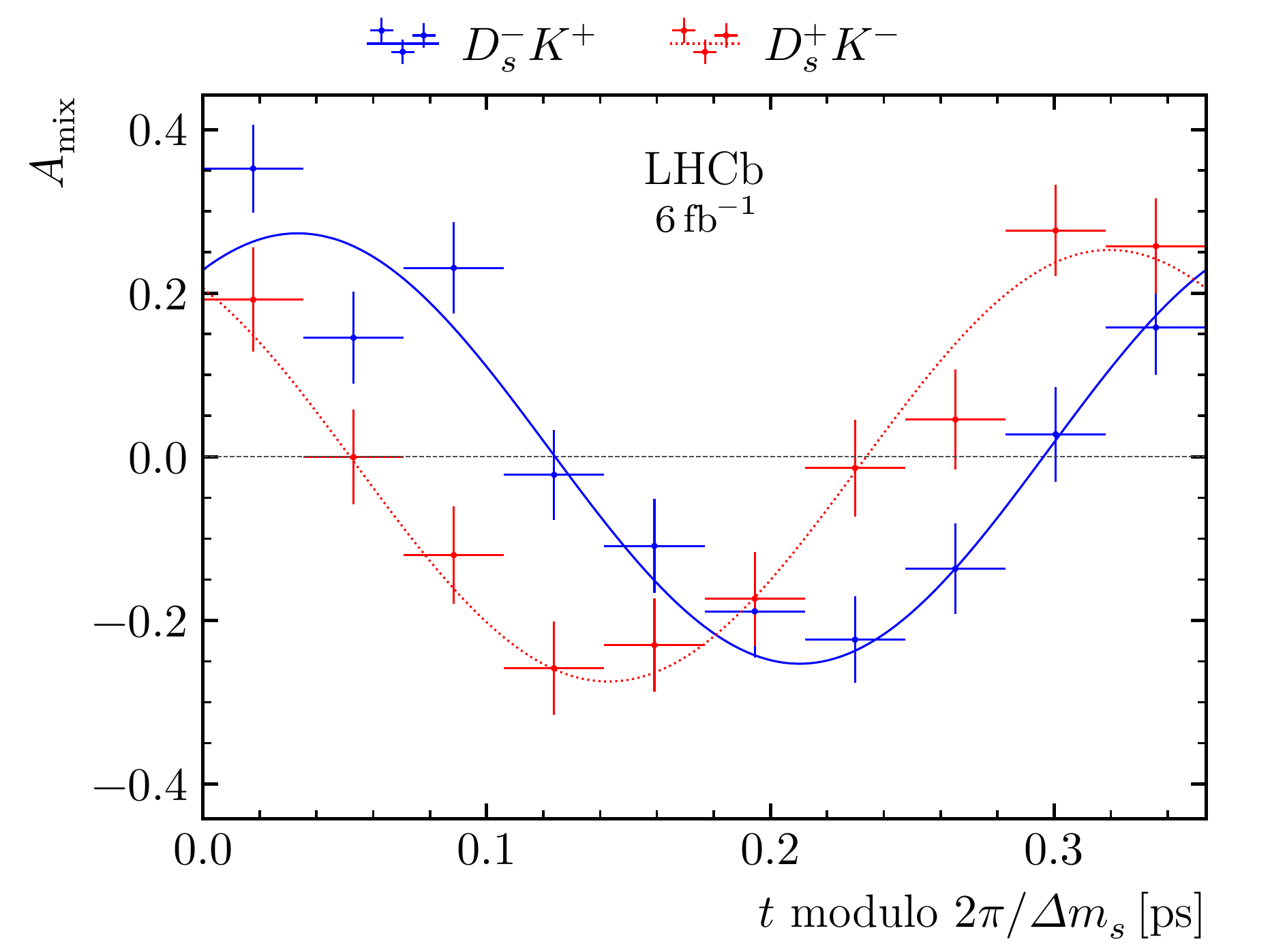}
	\caption{(Top) Decay-time distribution of \BsDsK signal candidates, where the background is statistically subtracted using the \emph{sPlot} technique.
    (Bottom) Mixing asymmetry, $A_{\text{mix}}$, for the (blue) $\Dsm K^+$ and the
    (red) $\Dsp K^-$ final states, folded into one mixing period, $2\pi/\dms$.
    In both plots, the curves show the result of the decay-time fit.}
 \label{fig:timesfit_bsdsk}
\end{figure}

\begin{figure}[!tb]
  \centering
  \includegraphics[width=.6\textwidth]{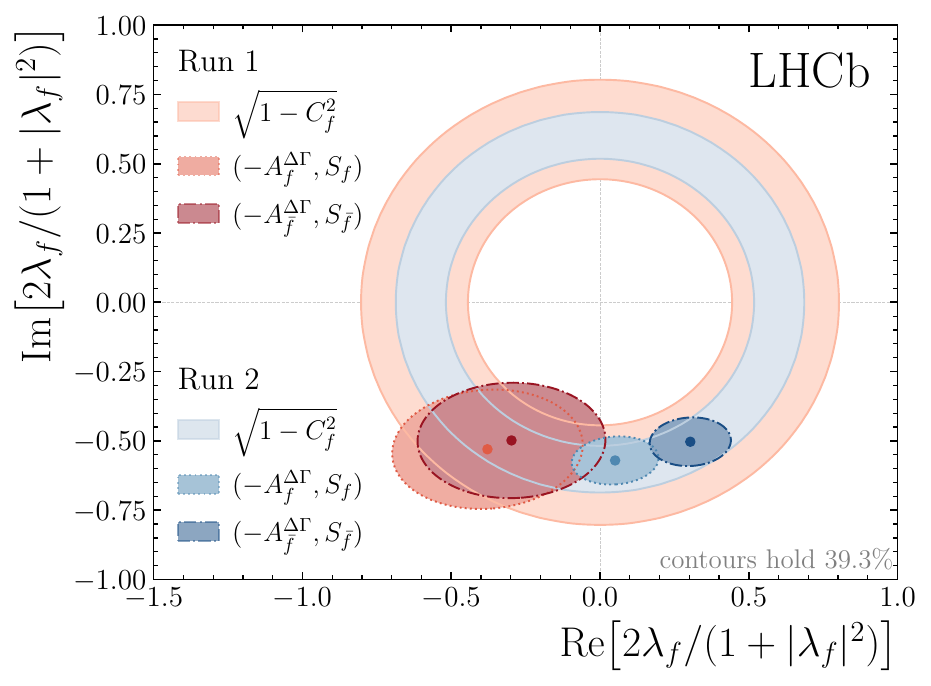}
\caption{
  \small Representation of the \CP observables in the (\mbox{$\Real\!\left[2\lambda_f/(1+|\lambda_f|^2)\right]$}, $\Imag\!\left[2\lambda_f/(1+|\lambda_f|^2)\right]$) Cartesian plane. The contours shown correspond to $39.3\%$ of the distribution.}
\label{fig:CPVparam-in-2d}
\end{figure}

The decay-time distribution and mixing asymmetry of the $\Dsm\Kp$ and $\Dsp\Km$ final states summed over all data-taking periods is shown in Fig.~\ref{fig:timesfit_bsdsk}, where the fit result is overlaid.
The mixing asymmetry is defined as the relative difference between events flavour-tagged as $\Bs$ and $\Bsb$, decaying to the $\Dsm\Kp$ or $\Dsp\Km$ final state as a function of decay time.
The \CP observables are represented in the ($\Real\!\left[2\lambda_f/(1+|\lambda_f|^2)\right]$, $\Imag\!\left[2\lambda_f/(1+|\lambda_f|^2)\right]$) Cartesian plane in Fig.~\ref{fig:CPVparam-in-2d}.
The agreement of the $(-\AdGamf, \Sf)$ and $(-\AdGamfb, \Sfb)$ contours with the $\sqrt{1-\Cf^2}$ band indicates that the results are in good agreement with the constraint $\Cf^2 + \Sf^2  + \AdGamf^2 = 1$ that
relates to Eq.~\ref{eq:truth}.
Using the statistical and systematic uncertainties reported in Table~\ref{tab:timefit_bsdsk} and the corresponding correlations, \CP violation in the interference of mixing and decay, \ie $\Sf \neq - \Sfb$, is observed with a significance of $8.6\,\sigma$.
The dependence of the \CP observables on the values of the \Gs, \DGs and \dms parameters is provided in Appendix~\ref{sec:GsDGsdms-dependence}.

\begin{table}[!t]
\centering
\caption{Statistical correlation matrix of the \CP observables.
Other fit parameters have negligible correlations with the \CP observables.
}
\label{tab:timefit_bsdskcorr}
\begin{tabular}{lrrrrr}
  \toprule
\rule{0pt}{2.5ex}    Parameter        & \Cf   & \AdGamf   & \AdGamfb  & \Sf   & \Sfb\\ 
  \midrule
 $\quad$ \Cf     & $\phantom{+}1 $  & $\phantom{+}0.134$ &$\phantom{+}0.130 $ & $\phantom{+}0.039$             & $\phantom{+}0.022 $  \\
 $\quad$ \AdGamf     &  & $\phantom{+}1 $  &$\phantom{+} 0.501 $  & $-0.108$             & $-0.036$  \\
 $\quad$ \AdGamfb    &  &  &$\phantom{+} 1 $  & $-0.056$             & $-0.067$  \\
 $\quad$ \Sf     &  &  &   & $\phantom{+}1 $  & $\phantom{+}0.006$  \\
 $\quad$ \Sfb    &  &  &   &  & $\phantom{+}1 $  \\
 \bottomrule
\end{tabular}
\end{table}

\section{Systematic uncertainties}
\label{sec:systematics}

Systematic uncertainties are evaluated due to various factors such as the modelling of the
invariant-mass fit, fixed parameters in the decay-time fit, namely those in Eq.~\ref{eq:decay-fit-inputs},
and the limited knowledge of the decay-time resolution and acceptance. Additionally,
the effect of ignoring correlations among observables is assessed.
Table~\ref{tab:TotalSystErrSfit} summarises the various contributions to the systematic uncertainty, expressed as fractions of the corresponding statistical uncertainties. These contributions are detailed below, with the relevant paragraphs beginning with the corresponding source in bold as listed in Table~\ref{tab:TotalSystErrSfit} for clarity.
The total systematic correlation matrix, reported in Table~\ref{tab:TotalSystCorSfit}, 
is obtained by adding the covariance matrices corresponding to each source.

\textbf{Invariant-mass fit} $\:$ The systematic uncertainty due to fixed parameters in the invariant-mass fit is determined by
repeating fits to the data, in which the fixed parameters for signal and background shapes are
varied by $\pm1$ standard deviation. 
Additionally, fixed background yields and relative fractions among background components are varied. 
For each parameter, the average difference of the \CP observables between the baseline and the modified fit is taken as the systematic uncertainty. 
After evaluating single contributions, all sources are added in quadrature.

\textbf{Flavour tagging} $\:$ The flavour-tagging parameters are constrained to the values found in the \BsDsPi decay-time fit.
Systematic uncertainties are assigned by performing a single \BsDsK decay-time fit, where the uncertainties on the flavour-tagging parameters are enlarged to account for their systematic effects.
This fit accounts for both the variation of the fit strategy used in the \BsDsPi data fit, as described in Ref.~\cite{LHCb-PAPER-2021-005}, and the portability of the flavour-tagging calibration to \BsDsK decays, studied in simulation.
This alternative decay-time fit is compared to the baseline fit and the difference in the results is assigned as a systematic uncertainty.

The remaining sources of systematic uncertainty are evaluated using pseudoexperiments, where each sample is generated according to the baseline results of the invariant-mass fit and, for the signal component, of the decay-time fit.
The decay-time distributions of the backgrounds are generated to resemble the properties observed in data or found in simulation.
The pseudoexperiments are processed using the same fit procedure as the data.
To assess the systematic uncertainty, the parameter in question, considered for the systematic effect, is varied within its uncertainty.
The values obtained from the baseline fit are compared to the values from the fits with the modified model. A distribution of the resulting differences is formed, taking into account the correlations between the baseline and modified pseudoexperiments. The systematic uncertainty is assigned as the mean and width of this distribution added in quadrature.

\textbf{Oscillation frequency $\Delta m_s$, Detection asymmetry $A_{\rm det}$} $\:$ The systematic uncertainty related to the uncertainty on \dms, which is fixed in the baseline fit,  is evaluated in a fit to pseudoexperiments in which the \dms value is
shifted by one standard deviation. The difference between the result of the modified fit and baseline value is assigned as the systematic uncertainty. In a similar way, the systematic uncertainty related to the uncertainty on the uncorrelated parameter $A_{\rm det}$ is determined.

\textbf{Decay-time resolution model} $\:$ The systematic uncertainty from the limited knowledge of the decay-time resolution is obtained by repeating the fit to pseudoexperiments using a narrower or wider decay-time resolution model. The largest deviation between the baseline and modified fits is taken as the systematic uncertainty. 

\textbf{Decay-time acceptance, \Gs, \DGs} $\:$ The correlation among the parameters \Gs, \DGs and the decay-time acceptance obtained from the \BsDsPi data necessitates a combined treatment of the corresponding systematic uncertainties.  
The pseudoexperiments are fitted with modified values of the spline coefficients, as well as the \Gs and \DGs parameters, which are sampled from the corresponding multidimensional correlated Gaussian distributions centred on their baseline values.
The combined correlated systematic uncertainty is determined from the average deviation of these modified fits with respect to the baseline fits.

\textbf{Decay-time acceptance simulation} $\:$ As the decay-time acceptance obtained from the \BsDsPi data is corrected by the simulation, an additional source of systematic uncertainty arises from the limited size of the simulation samples that are used to determine the correction. To account for this, pseudoexperiments are fitted with a modified decay-time acceptance correction, that is sampled from a multidimensional Gaussian representing the uncertainties and correlations of the corrections. The average deviation of these modified fits with respect to the baseline fits is taken as a systematic uncertainty.

\textbf{Decay-time bias} $\:$ A decay-time bias is observed in the \dms measurement from the \BsDsPi decays~\cite{LHCb-PAPER-2021-005}, and corrected for in the baseline fit.
A systematic uncertainty related to the uncertainty on the bias, taken to be $\pm 1\fs$, is obtained using pseudoexperiments.
This value includes both the uncertainty of the decay-time bias as obtained in the \BsDsPi decays and the correction to account for the differences in selection criteria between the two analyses~\cite{LHCb-PAPER-2021-005}.

\begin{table}[!tb]
\centering
\caption{Systematic uncertainties on the \CP observables, expressed as a fraction of the corresponding statistical uncertainties. The value ``---" indicates that the contribution is negligible.} 
\label{tab:TotalSystErrSfit}
\begin{tabular}{lccccc}
\toprule
Source                                   & \Cf  & 
\AdGamf   & \AdGamfb & \Sf  & \Sfb \\
\midrule
Invariant-mass fit                 & 0.045 & 0.095 & 0.121 & 0.088 & 0.112 \\
Flavour tagging                  & 0.256 & 0.026 & 0.028 & 0.012 & 0.070 \\
Oscillation frequency $\Delta m_s$                     & 0.006 & 0.005 & 0.004 & 0.108 & 0.101 \\
Detection asymmetry $A_{\rm det}$              & 0.001 & 0.079 & 0.082 & 0.007 & 0.007 \\
Decay-time resolution model      & 0.195 & 0.008 & 0.008 & 0.054 & 0.166 \\
Decay-time acceptance, \Gs, \DGs & 0.006 & 0.397 & 0.400 & 0.009 & 0.009 \\
Decay-time acceptance simulation    & 0.004 & 0.064 & 0.064 & --- & 0.004 \\ 
Decay-time bias                  & 0.062 & 0.027 & 0.046 & 0.188 & 0.167 \\
Neglecting correlations          & 0.137 & 0.081 & 0.054 & 0.135 & 0.043 \\
\midrule
Total                            & 0.358 & 0.430 & 0.439 & 0.277 & 0.293 \\
\bottomrule
\end{tabular}
\end{table}

\textbf{Neglecting correlations} $\:$ The impact of neglecting the correlations among decay time and decay-time uncertainty with the \Bs and \Ds invariant masses in the \emph{sPlot} method is studied with a dedicated set of pseudoexperiments using a bootstrapping method~\cite{efron:1979}.
The method preserves correlations among observables. The results of the decay-time fits performed on samples with and without these correlations are compared and the mean difference is assigned as the systematic uncertainty.

Additional cross-checks are performed to further validate the results. 
The data sample is split into subsets according to the 
two LHCb dipole magnet polarity orientations, the year of data taking and the \Bs meson momentum. In addition, the data sample is split into decay-time bins and the invariant-mass fit is performed on each subsample. This is followed by the combined decay-time fit.  
In all cases, no significant deviations among the results are observed. 
A closure test using a large sample of simulated signal candidates provides an estimate 
of the intrinsic bias related to the decay-time fit procedure. 
No significant bias is found. 

Several other potential systematic effects are examined but are found to be
negligible.
No significant systematic effect associated with the production asymmetry $A_{\rm prod}$ is observed. 
The decay-time fit is repeated by varying the knot positions in the decay-time acceptance
description.
No significant changes with respect to the baseline result are found. 
The precision on the world-average value for the oscillation frequency \dms is dominated by the LHCb measurement~\cite{LHCb-PAPER-2021-005}, which uses the same \BsDsPi sample used as a control channel in this analysis.
The imperfect knowledge of the particles' momentum and the longitudinal dimension of the
detector are encompassed by the systematic uncertainty on \dms, therefore these sources
are not further considered. 

\begin{table}[tb]
\centering
\caption{Correlation matrix of the total systematic uncertainties of the \CP observables.}
\label{tab:TotalSystCorSfit}
\begin{tabular}{crrrrr}
\toprule
Parameter & \Cf  & \AdGamf   & \AdGamfb & \Sf  & \Sfb \\
\midrule
 $\quad$ \Cf      & $1$ & $0.008$ & $0.012$ & $-0.080$           & $-0.246$  \\
 $\quad$ \AdGamf  &     & $1$     & $0.878$ & $\phantom{+}0.004$ & $-0.022$  \\
 $\quad$ \AdGamfb &     &         & $1$     & $-0.002$           & $-0.022$  \\
 $\quad$ \Sf      &     &         &         & $\phantom{+}1 $    & $\phantom{+}0.085$  \\
 $\quad$ \Sfb     &     &         &         &                    & $\phantom{+}1 $  \\
\bottomrule
\end{tabular}
\end{table}

\section{Interpretation}
\label{sec:interpretation}

The measured $\CP$-violating parameters listed in Table~\ref{tab:timefit_bsdsk} can be interpreted in terms of the CKM angle $\gamma$, the strong phase difference $\delta$, the magnitude of the amplitude ratio $\rdsk$ and the mixing phase $\beta_s$. 
This is achieved using a frequentist approach detailed in Refs.~\cite{GammaCombo, LHCb-PAPER-2016-032}.
First, a likelihood function is defined as
\begin{equation}
    \mathcal{L}(\vec{\alpha}) = \prod_{i} f (\vec{A}_i^{\rm{obs}}|\vec{\alpha}) \ ,
\end{equation}
where the function  $f (\vec{A}_i^{\rm{obs}}|\vec{\alpha})$ is assumed to follow a multivariate
Gaussian distribution
\begin{equation}
    f(\vec{A}_i^{\rm{obs}}|\vec{\alpha}) \propto \exp \left( -\frac{1}{2} ( \vec A(\vec \alpha) - \vec{A}^{\rm{obs}}_{i} )^T \,V_i^{-1} \, ( \vec{A}(\vec \alpha) - \vec{A}^{\text{obs}} ) \right) \ ,
\end{equation}
where $\vec{A}_i^{\rm{obs}}$ is the vector of observables $(\Cf, \AdGamf, \AdGamfb, \Sf, \Sfb)$, $V_i$ is the experimental covariance matrix, and $i$ is an index labelling the different measurements to be combined.
The vector function $\vec{A}(\vec{\alpha})$ encodes the dependency of the observables on the parameters of interest $\vec{\alpha} = (\gamma, \beta_s, \delta, \rdsk)$, following Eq.~\ref{eq:truth}.
A fit is performed to find the set of values $\vec{\alpha}_\text{min}$ that minimise the function 
$\chi^2(\vec{\alpha}) = -2 \ln \mathcal{L}(\vec{\alpha})$.
An ensemble of pseudoexperiments is generated to determine the best-fit values and confidence intervals of the parameters $\vec{\alpha}$, as detailed in Ref.~\cite{Bodhisattva:2009uba}.
This method is referred to as the \textit{Plugin} method and is used throughout this paper for the results of $\gamma$, $\delta$ and $\rdsk$.

As discussed in Sec.~\ref{sec:Introduction}, the \CP observables determined from
\BsDsK decays are functions of the weak-phase difference $(\gamma -2 \beta_s)$.
Therefore, in order to determine $\gamma$, the value of $\beta_s$ has to be taken from independent measurements.
The value of $\beta_s$ is obtained through the relation $\phi_s = -2 \beta_s $.
Neglecting the contributions from loop diagrams, whose impact is estimated to be below the statistical uncertainty on $\phis$~\cite{Barel:2020jvf}, the value $\phis = -0.031 \pm 0.018 \rad$ is used, which is taken from the combination of LHCb measurements presented in Ref.~\cite{LHCb-PAPER-2023-016}.

\begin{figure}
  \centering
  \includegraphics[width=.48\textwidth]{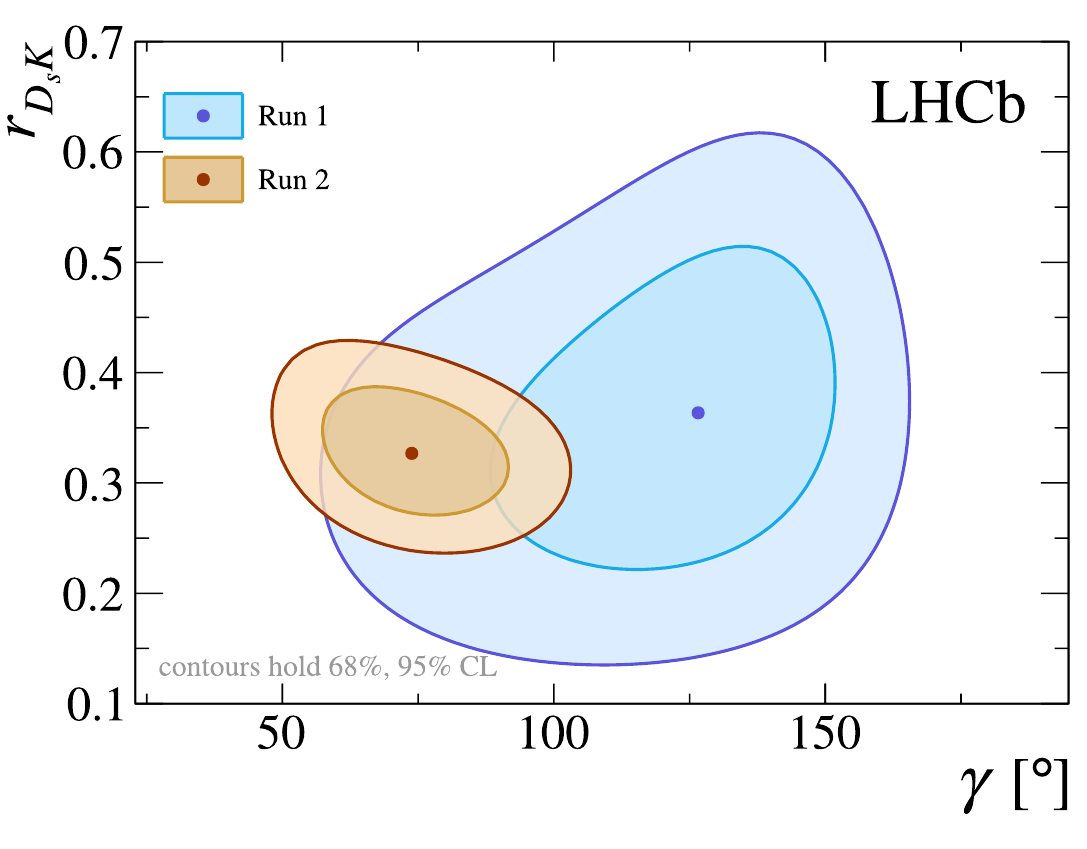}
  \includegraphics[width=.48\textwidth]{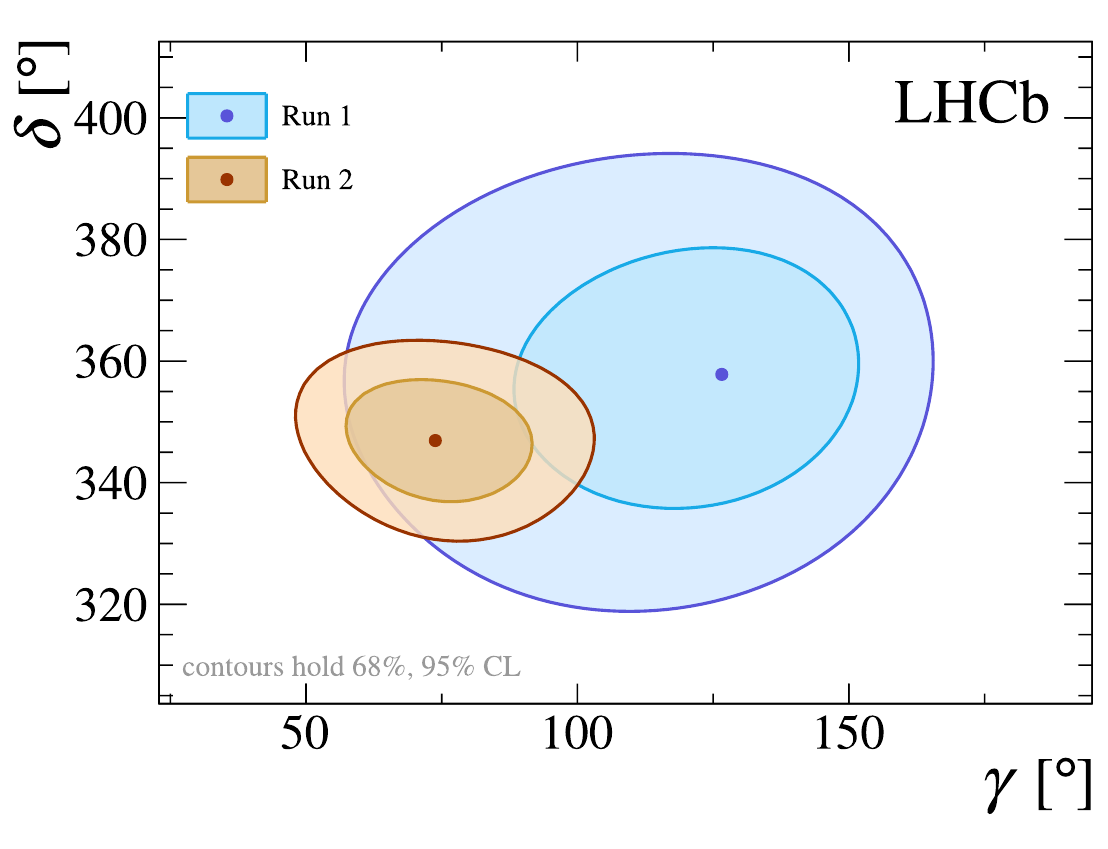} \\
  \includegraphics[width=.55\textwidth]{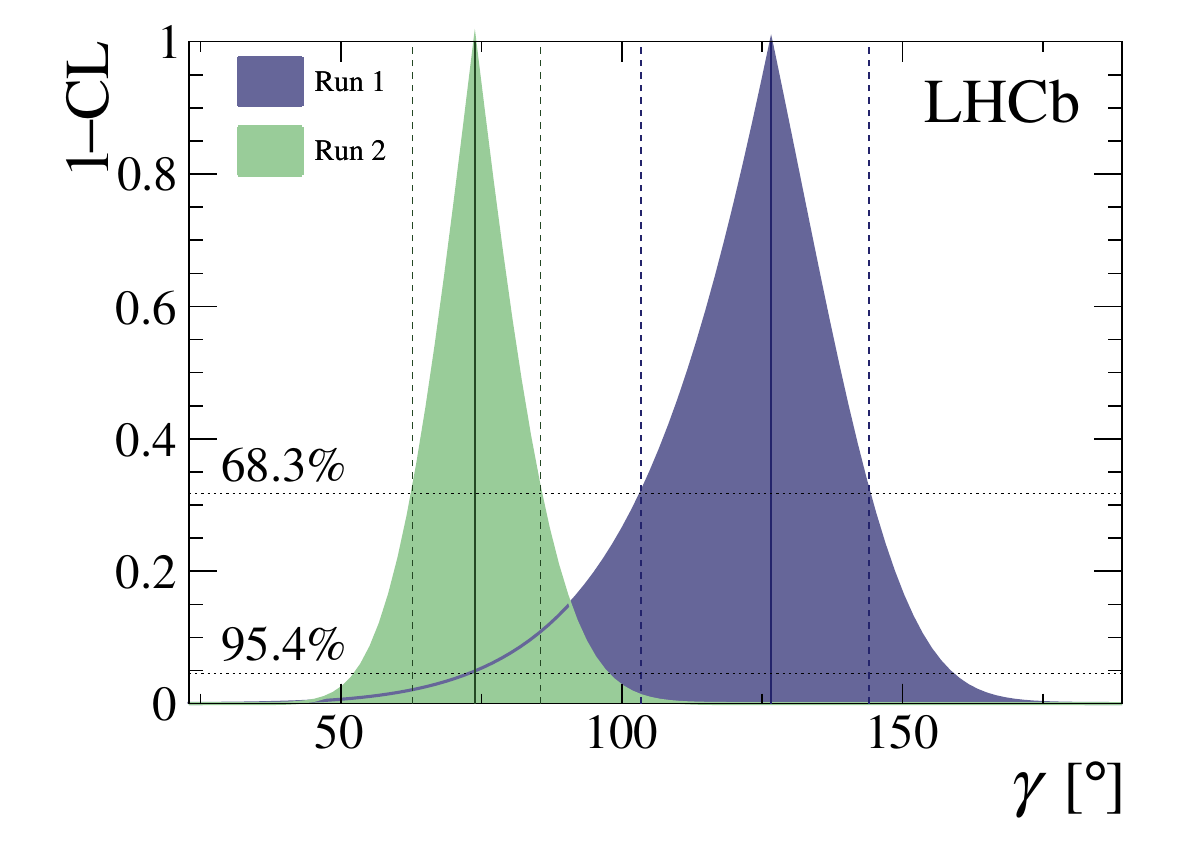} 
\caption{
  \small
  Contour plots of (top left) \rdsk \vs~\g and (top right)
  \strong \vs~\g, where the contours correspond to the confidence levels (CL) of 68\% and 95\%.
  The bottom plot shows the \omcl curve for the angle \g, with the 68.3\% and 95.4\% CL intervals indicated with horizontal and vertical lines.
  The results correspond to $\g = (74\pm12)^\circ$ and $\g =  ( 127^{+18}_{-26} )^\circ$ for the Run 2 and Run 1 analysis, respectively,  where updated values of \Gs, \DGs and \dms are also propagated to the
  latter.
}
\label{fig:interpretation_gamma}
\end{figure}

The $\CP$-violating parameters obtained from the Run~2 $\BsDsK$ data correspond to the following parameters:
\begin{align*}
    \g      &=  (74\pm12)^\circ\ ,\\
    \strong &= (346.9^{+6.8}_{-6.6})^\circ\ ,\\
    \rdsk   &= 0.327^{+0.039}_{-0.037}\ ,
\end{align*}
where the phases $\g$ and $\strong$ are determined up to a global shift of $180^\circ$. The statistical and systematic uncertainties are summed in quadrature.
Figure~\ref{fig:interpretation_gamma} shows the confidence-level (CL) plot for $\gamma$, as well as the two-dimensional contours of $\gamma$ versus $\rdsk$ and $\delta$.

\section{Combination with the results of the Run~1 analysis}
\label{sec:combination}
The results of this measurement are combined with those obtained using Run~1 data, presented in Ref.~\cite{LHCb-PAPER-2017-047}.
To achieve that, the main systematic uncertainties of the Run~1 measurement related to the external parameters  \Gs, \DGs and \dms are recomputed using the more precise values considered in this analysis.
These external parameters only affect the time-dependent part of the analysis, which is therefore updated with the new values.
The following steps are repeated:
the determination of the decay-time acceptance from \BsDsPi data,
the flavour-tagging calibration, which is required because of its correlation with the \dms parameter in the fit to \BsDsPi data and
the decay-time fit to \BsDsK candidates to extract the values of the \CP observables.
Other aspects of the analysis, such as the decay-time resolution, obtained from calibration data, are unchanged with respect to Ref.~\cite{LHCb-PAPER-2017-047}.
The updated values of the \CP observables from the Run~1 data set are reported
in Table~\ref{tab:timefit_bsdsk_run1_updated} and correspond to $\CP$ violation with a significance of $3.4\,\sigma$.
\begin{table}[!bt]
\centering
\caption{Updated values of the \CP observables from the decay-time fit of the Run~1 analysis with updated values of the nuisance parameters \dms, \Gs and \DGs.
The first uncertainty is statistical and the second is systematic. }
\label{tab:timefit_bsdsk_run1_updated}
\begin{tabular}{lc}
  \toprule
  Parameter     & Value \\
  \midrule
  $\quad$ \Cf         & $\phantom{+}0.75 \pm  0.14 \pm 0.04$  \\ 
  $\quad$ \AdGamf     & $\phantom{+}0.38  \pm  0.28 \pm 0.15$  \\
  $\quad$ \AdGamfb    & $\phantom{+}0.30  \pm  0.28 \pm 0.15$  \\
  $\quad$ \Sf         & $ -0.53 \pm 0.21 \pm 0.06$  \\
  $\quad$ \Sfb        & $ -0.50 \pm 0.20 \pm 0.06$  \\
  \bottomrule
\end{tabular}
\vspace{-3mm}
\end{table}
Following the same procedure as described in Sec.~\ref{sec:interpretation}, the values of the
parameters $\gamma$, \strong and \rdsk corresponding to the Run~1 measurement are
\begin{align*}
    \g      &=  ( 127^{+18}_{-26} )^\circ\ ,\\
    \strong &= (358^{+14}_{-15})^\circ\ ,\\
    \rdsk   &= 0.364^{+0.095}_{-0.094}\ .
\end{align*}
The Run~2 and the updated Run~1 results are compared in Fig.~\ref{fig:interpretation_gamma}.
The compatibility of $\gamma$, \strong and \rdsk parameters between the updated Run~1 and the Run~2 results corresponds to a $p$-value of $12\%$.

\begin{figure}[!tb]
    \centering
    \includegraphics[width=0.48\linewidth]{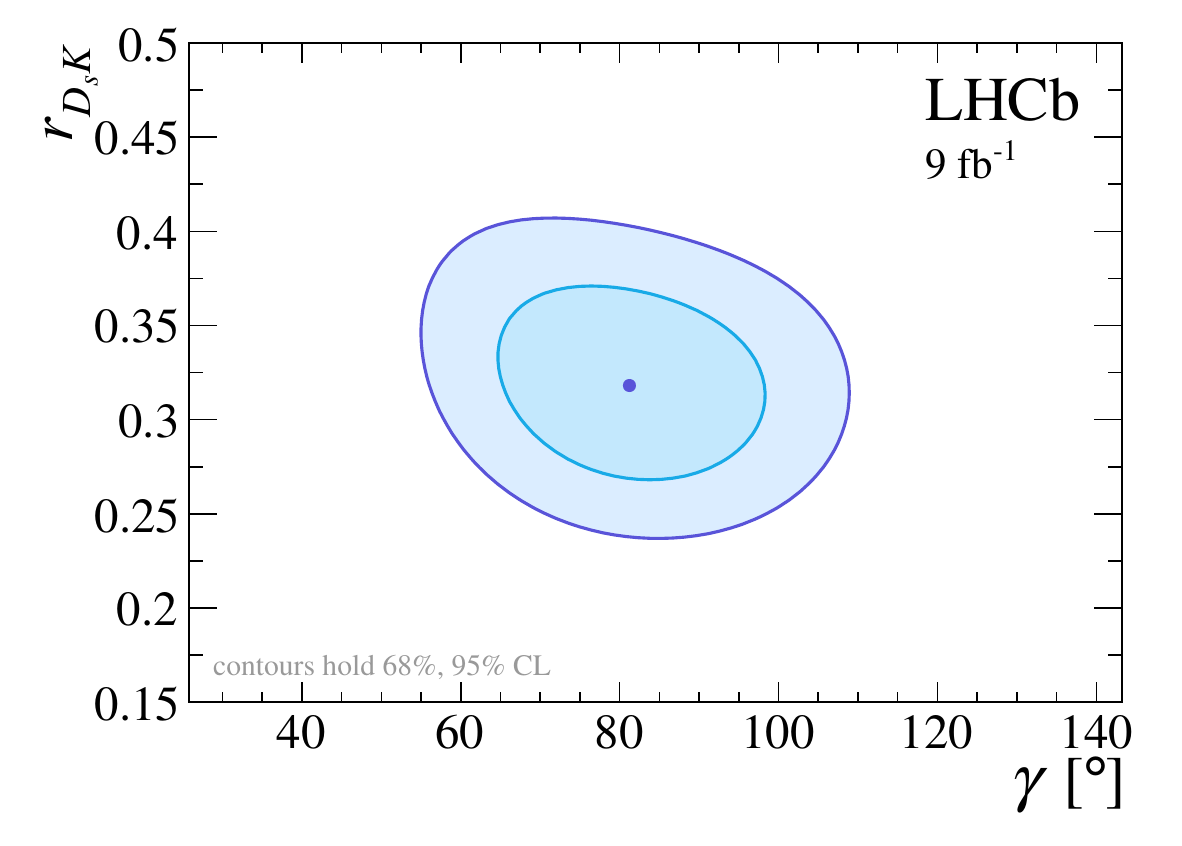}
    \includegraphics[width=0.48\linewidth]{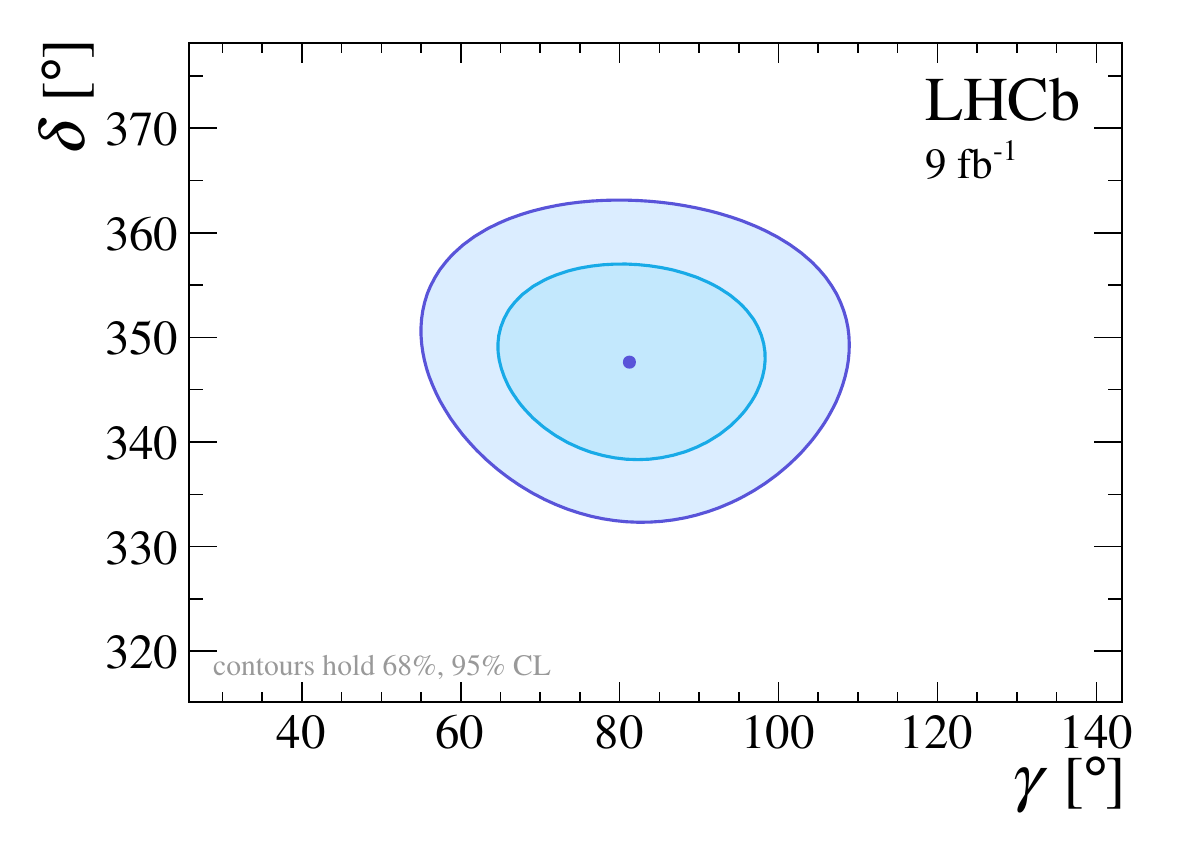}\\
    \includegraphics[width=0.55\linewidth]{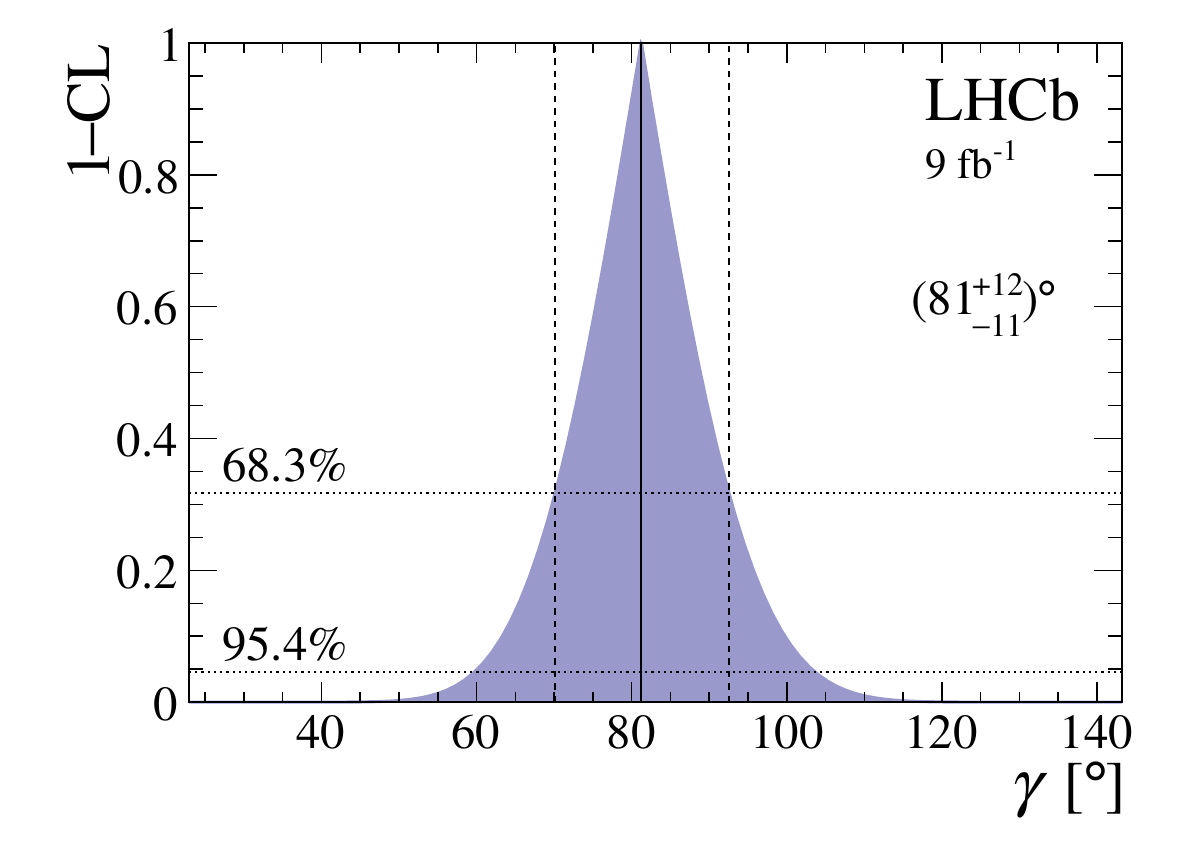}
    \caption{
        Contour plots arising from the combined extraction of $\gamma$, $\delta$ and $\rdsk$ from the \CP parameters from the Run~1 and Run~2 data sets. (Top left) \rdsk \vs~\g and (top right)
        \strong \vs~\g, the contours correspond to the confidence levels (CL) of 68\% and 95\%.
        The bottom plot shows the \omcl curve for the angle \g, with the 68.3\% and 95.4\% CL intervals indicated with horizontal and vertical lines.
    }
    \label{fig:run1_run2_combination_gamma}
\end{figure}

The \CP-violating parameters of the two data sets have been combined following the methods presented in Ref.~\cite{LHCb-CONF-2024-004,*LHCb-PAPER-2021-033}.
The two measurements are treated as independent and the optimal values of $\gamma$, $\strong$ and $\rdsk$ are obtained via the likelihood fit described in Sec.~\ref{sec:interpretation}.
The resulting values are
\begin{align*}
    \g      &=  ( 81^{+12}_{-11} )^\circ\ ,\\
    \strong &= (347.6\pm 6.3)^\circ\ ,\\
    \rdsk   &= 0.318^{+0.034}_{-0.033}\ .
\end{align*}
The corresponding $1-\textrm{CL}$ curve for \g is shown in
Fig.~\ref{fig:run1_run2_combination_gamma}, as well as the two-dimensional contours of $\gamma$ versus $\rdsk$ and $\delta$.
Finally, the value of the relative weak-phase difference in $\BsDsK$ decays is determined to be $\gamma -2 \beta_s = (79^{+12}_{-11})^\circ$, providing complementary sensitivity to $\phi_s$ on a potential new physics phase in $\Bs$--$\Bsb$ mixing.

The biggest contribution to the increased sensitivity to $\gamma$ is the improved statistical power of the Run 2 sample with respect to Run 1. 
This improvement results from the increase of integrated luminosity, $\bquark \bquarkbar$ production cross section, trigger and signal selection efficiencies and an improved flavour tagging performance.

\section{Conclusion}
\label{sec:conclusion}

The \CP-violating parameters that describe the \BsDsK decay rates
have been measured using a data set corresponding to an integrated luminosity of $6 \invfb$ of $pp$ collisions recorded with the \lhcb detector.
Their values are found to be
\begin{align*}
\Cf         &= \phantom{+}0.791  \pm  0.061 \pm 0.022\ ,  \\ 
\AdGamf     &= -0.051  \pm 0.134 \pm 0.058\ ,  \\
\AdGamfb    &= -0.303  \pm 0.125 \pm 0.055\ ,  \\
\Sf         &= -0.571  \pm 0.084 \pm 0.023\ ,  \\
\Sfb        &= -0.503  \pm 0.084 \pm 0.025\ ,  
\end{align*}
where the first uncertainties are statistical and the second are systematic.
$\CP$ violation in the interference between $\Bs$--$\Bsb$ mixing and $\BsDsK$ decays is observed with a significance of $8.6\,\sigma$.
The results are used to determine the CKM angle \g, the strong-phase difference $\delta$ and 
the magnitude of the ratio $\rdsk$ between the 
\BsDspKm and the \BsDsmKp decay
amplitudes, leading to
\begin{align*}
    \g      &=  (74\pm12)^\circ\ ,\\
    \strong &= (346.9^{+6.8}_{-6.6})^\circ\ ,\\
    \rdsk   &= 0.327^{+0.039}_{-0.037}\ ,
\end{align*}
where all angles are given modulo~$180^\circ$, and uncertainties shown are the combination of the 
statistical and systematic contributions.

The results of the present analysis are combined with those from the previous \lhcb 
analysis~\cite{LHCb-PAPER-2017-047}, which is updated to account for improved determinations of \Gs, \DGs and 
\dms values.
The following values of \g, $\delta$ and $\rdsk$ are found from the combination:
\begin{align*}
    \g      &=  ( 81^{+12}_{-11} )^\circ\ ,\\
    \strong &= (347.6\pm 6.3)^\circ\ ,\\
    \rdsk   &= 0.318^{+0.034}_{-0.033}\ .
\end{align*}
This result represents the most precise determination of $\g$ in $\Bs$ meson decays and is in good agreement, within $1.4\, \sigma$, with the most recent \lhcb combination~\cite{LHCb-CONF-2024-004,*LHCb-PAPER-2021-033}.

\appendix 

\section*{Acknowledgements}
%
%
\noindent We express our gratitude to our colleagues in the CERN
accelerator departments for the excellent performance of the LHC. We
thank the technical and administrative staff at the LHCb
institutes.
We acknowledge support from CERN and from the national agencies:
CAPES, CNPq, FAPERJ and FINEP (Brazil); 
MOST and NSFC (China); 
CNRS/IN2P3 (France); 
BMBF, DFG and MPG (Germany); 
INFN (Italy); 
NWO (Netherlands); 
MNiSW and NCN (Poland); 
MCID/IFA (Romania); 
MICIU and AEI (Spain);
SNSF and SER (Switzerland); 
NASU (Ukraine); 
STFC (United Kingdom); 
DOE NP and NSF (USA).
We acknowledge the computing resources that are provided by CERN, IN2P3
(France), KIT and DESY (Germany), INFN (Italy), SURF (Netherlands),
PIC (Spain), GridPP (United Kingdom), 
CSCS (Switzerland), IFIN-HH (Romania), CBPF (Brazil),
and Polish WLCG (Poland).
We are indebted to the communities behind the multiple open-source
software packages on which we depend.
Individual groups or members have received support from
ARC and ARDC (Australia);
Key Research Program of Frontier Sciences of CAS, CAS PIFI, CAS CCEPP, 
Fundamental Research Funds for the Central Universities, 
and Sci. \& Tech. Program of Guangzhou (China);
Minciencias (Colombia);
EPLANET, Marie Sk\l{}odowska-Curie Actions, ERC and NextGenerationEU (European Union);
A*MIDEX, ANR, IPhU and Labex P2IO, and R\'{e}gion Auvergne-Rh\^{o}ne-Alpes (France);
AvH Foundation (Germany);
ICSC (Italy); 
Severo Ochoa and Mar\'ia de Maeztu Units of Excellence, GVA, XuntaGal, GENCAT, InTalent-Inditex and Prog. ~Atracci\'on Talento CM (Spain);
SRC (Sweden);
the Leverhulme Trust, the Royal Society
 and UKRI (United Kingdom).

\clearpage

\section*{Appendices}

\appendix

\section{Dependence of the \texorpdfstring{$\boldsymbol{\CP}$}{CP}-violating parameters on external parameters for Run~2 data}
\label{sec:GsDGsdms-dependence}

The \CP-violating parameters depend on external parameters such as the mixing frequency, \dms, the \Bs decay width, \Gs, and the decay-width difference, \DGs, which are fixed in the baseline \BsDsK decay-time fit.
The central values of these parameters might evolve over years, thus for any future combinations of the CKM angle $\gamma$ it is important to evaluate the shift of the \CP-violating parameters as a function of external parameters.
This dependence on external parameters is calculated separately for Run~1 and Run~2 data-taking periods.

\begin{figure}[!tb]
  \centering
  \includegraphics[width=.75\textwidth]{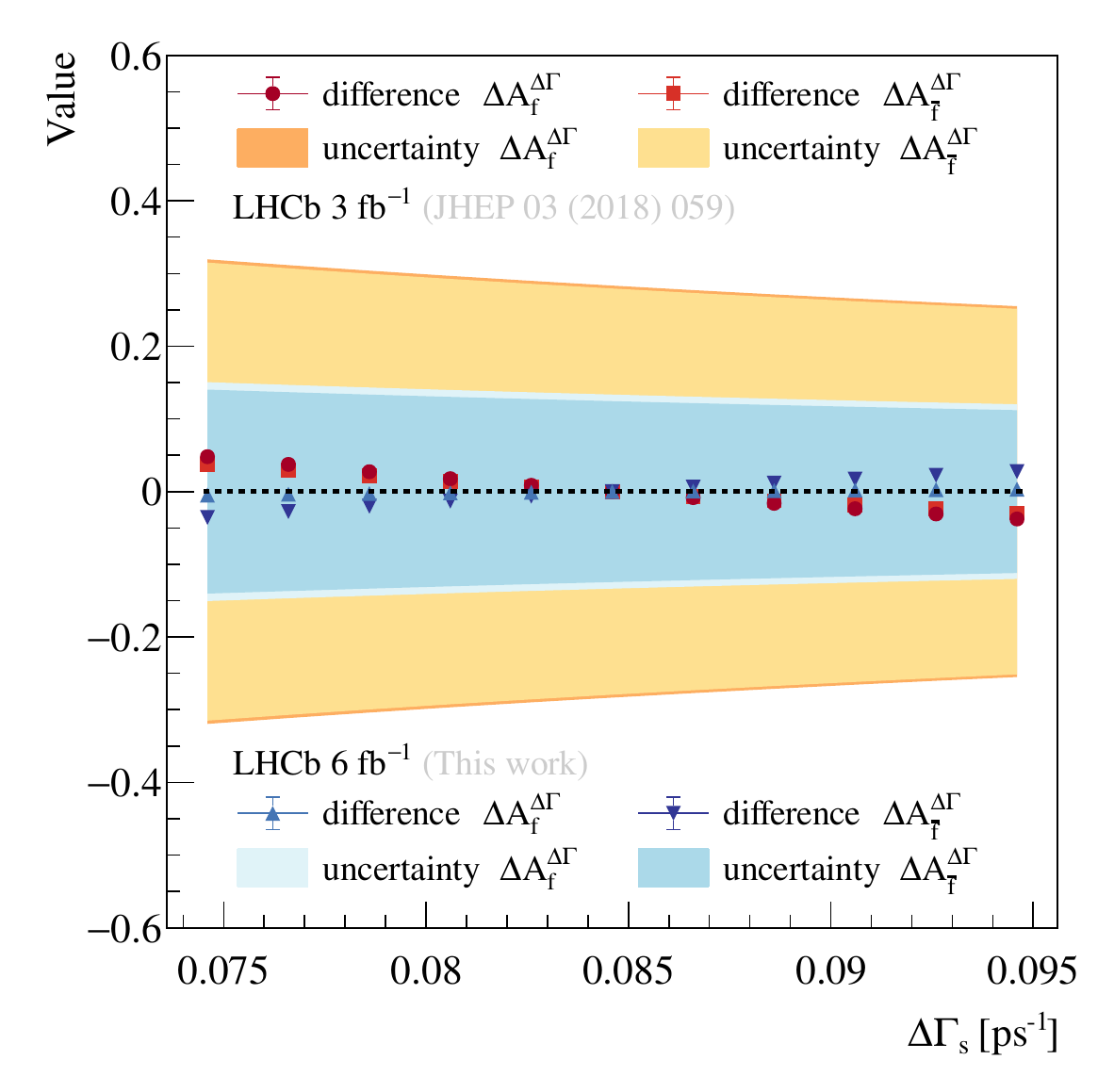}
  \caption{
    Dependence of the \CP-violating parameters \AdGamf and \AdGamfb on the \DGs parameter.
The triangles, circles and squares denote the differences in the \AdGamf and \AdGamfb parameters with respect to the baseline result. 
For comparison, the orange and blue bands visualise the size of the statistical uncertainty of the \AdGamf and \AdGamfb parameters as obtained in the corresponding fits to Run~1 and Run~2 data, respectively.
  }
  \label{fig:Df-Dfb-vs-DGs}
\end{figure}

The dependence of the \CP-violating parameters \AdGamf and \AdGamfb on \DGs
is shown in Fig.~\ref{fig:Df-Dfb-vs-DGs} in the interval
\DGs $\in$ [0.075, 0.095] ps$^{-1}$. 
The horizontally and vertically hatched bands represent the
statistical uncertainty of the \AdGamf and \AdGamfb parameters,
while circles (squares) denote the difference of the \AdGamf
(\AdGamfb) parameter with respect to the baseline result.
The interval is extended with respect to the baseline range, \DGs $\in$ [0.081, 0.089] ps$^{-1}$, to account for the differences in the \DGs determination
obtained with and without constraints from effective lifetime measurements~\cite{HFLAV21}. 
The decay-time fit is repeated for ten alternative \DGs values within the extended interval. For these alternative fits the decay width, \Gs, is fixed to be \mbox{$\Gs= 0.6563$ \invps}. The difference of the \CP-violating parameters with respect to the baseline result is evaluated. A small dependence is observed for the \AdGamf and \AdGamfb parameters.
The first derivatives of the \AdGamf and \AdGamfb parameters with respect to the \DGs variable
are determined to be $0.400 \pm 0.012\ps$ and $3.13 \pm 0.07\ps$, respectively, for Run~2.
For Run~1, the corresponding values are $-4.26 \pm 0.10\ps$ and $-3.34 \pm 0.08\ps$.
The derivatives are found to be negligible for the  \Cf, \Sf and \Sfb parameters.
Similarly, the dependence on the \Gs parameter is studied and found to be negligible for all \CP-violating observables.

The dependence of the \CP-violating parameters on the mixing frequency, \dms, is obtained from pseudoexperiments, where the \dms value 
is shifted by one standard deviation with respect to the baseline value. In these alternative fits, the \DGs parameter is fixed to be $\DGs=0.085 \invps$.
The first derivatives of \Sf and \Sfb computed with respect to the \dms variable
are evaluated to be $1.712 \pm 0.011 \ps$ and $-1.653 \pm 0.011\ps$, respectively, for Run~2.
For Run~1, the corresponding values are $1.6\pm0.9 \ps$ and $-1.5\pm1.0 \ps$.
The dependencies are found to be negligible for the \Cf, \AdGamf and \AdGamfb parameters.

\section{Total statistical and systematic correlation matrices for the updated Run~1 result}

For the combination of Run~1 and Run~2 results described in Sec.~\ref{sec:combination} the decay-time fit of the Run 1 measurement is updated with more recent values of input parameters used in the present analysis of Run~2 data. For reference, the correlation matrix of the five parameters extracted in the updated fit is given in Table~\ref{tab:timefit_bsdskcorr_run1_updated}.
Additionally, the update implies the need to reevaluate some of the systematic uncertainties.
The systematic uncertainties in the Run 1 measurement introduced by the limited knowledge of the parameters \dms, \Gs, \DGs and the decay-time acceptance model are evaluated with the same pseudoexperiment-based approach as described in Sec.~\ref{sec:systematics}.
The total systematic uncertainty, with updated contributions, is reported together with the updated statistical uncertainty in Table~\ref{tab:timefit_bsdsk_run1_updated}.
For completeness, the correlation matrix of the combined and partially updated systematic uncertainty of the Run 1 measurement is provided in Table~\ref{tab:TotalSystCorSfit_run1}.

\begin{table}[t]
\centering
\caption{Statistical correlation matrix of the \CP observables in the updated fit of the Run~1 data set.}
\label{tab:timefit_bsdskcorr_run1_updated}
\begin{tabular}{lrrrrr}
  \toprule
Parameter        & \Cf   & \AdGamf   & \AdGamfb  & \Sf   & \Sfb\\ 
  \midrule
 $\quad$ \Cf      & $1$ & $0.114$ & $0.098$ & $0.018$ & $-0.054$ \\
 $\quad$ \AdGamf  & & $1$ & $0.546$ & $-0.088$ & $-0.024$ \\
 $\quad$ \AdGamfb & & & $1$ & $-0.051$ & $-0.024$ \\
 $\quad$ \Sf      & & & & $1$ & $0.001$ \\
 $\quad$ \Sfb     & & & & & $1$ \\
 \bottomrule
\end{tabular}
\end{table}

\begin{table}[htb!]
\centering
\caption{Systematic correlation matrix of the \CP observables in the updated fit of the Run~1 data set.}
\label{tab:TotalSystCorSfit_run1}
\begin{tabular}{crrrrr}
\toprule
Parameter & \Cf  & \AdGamf   & \AdGamfb & \Sf  & \Sfb \\
\midrule
 $\quad$ \Cf      & $1$ & $0.07$ & $0.05$ & $0.04$ &$-0.01$ \\
 $\quad$ \AdGamf  & & $1$ & $0.53$ & $0.02$ & $0.02$ \\
 $\quad$ \AdGamfb & & & $1$ & $0.03$ & $0.03$ \\
 $\quad$ \Sf      & & & & $1$ & $0.02$ \\
 $\quad$ \Sfb     & & & & & $1$ \\
\bottomrule
\end{tabular}
\end{table}




\addcontentsline{toc}{section}{References}
\bibliographystyle{LHCb}
\bibliography{main,standard,LHCb-PAPER,LHCb-CONF,LHCb-DP,LHCb-TDR}

\newpage
\centerline
{\large\bf LHCb collaboration}
\begin
{flushleft}
\small
R.~Aaij$^{36}$\lhcborcid{0000-0003-0533-1952},
A.S.W.~Abdelmotteleb$^{55}$\lhcborcid{0000-0001-7905-0542},
C.~Abellan~Beteta$^{49}$,
F.~Abudin{\'e}n$^{55}$\lhcborcid{0000-0002-6737-3528},
T.~Ackernley$^{59}$\lhcborcid{0000-0002-5951-3498},
A. A. ~Adefisoye$^{67}$\lhcborcid{0000-0003-2448-1550},
B.~Adeva$^{45}$\lhcborcid{0000-0001-9756-3712},
M.~Adinolfi$^{53}$\lhcborcid{0000-0002-1326-1264},
P.~Adlarson$^{80}$\lhcborcid{0000-0001-6280-3851},
C.~Agapopoulou$^{13}$\lhcborcid{0000-0002-2368-0147}
% ...

{\footnotesize \it

% ...
$^{13}$Universit{\'e} Paris-Saclay, CNRS/IN2P3, IJCLab, Orsay, France\\
% ...
$^{36}$Nikhef National Institute for Subatomic Physics, Amsterdam, Netherlands\\
% ...
$^{45}$Instituto Galego de F{\'\i}sica de Altas Enerx{\'\i}as (IGFAE), Universidade de Santiago de Compostela, Santiago de Compostela, Spain\\
% ...
$^{49}$Physik-Institut, Universit{\"a}t Z{\"u}rich, Z{\"u}rich, Switzerland\\
% ...
$^{53}$H.H. Wills Physics Laboratory, University of Bristol, Bristol, United Kingdom\\
% ...
$^{55}$Department of Physics, University of Warwick, Coventry, United Kingdom\\
% ...
$^{58}$School of Physics and Astronomy, University of Glasgow, Glasgow, United Kingdom\\
$^{59}$Oliver Lodge Laboratory, University of Liverpool, Liverpool, United Kingdom\\
% ...
$^{67}$Syracuse University, Syracuse, NY, United States\\
% ...
$^{80}$Department of Physics and Astronomy, Uppsala University, Uppsala, Sweden, associated to $^{58}$\\
% ...
}
\end{flushleft}

\end{document}